\documentstyle[pre,aps,eqsecnum,epsf]{revtex}

\newcommand{\mean}[1]{\left\langle \, {#1} \, \right\rangle}

\newcommand{\locav}[1]{\left\lceil{#1}\right\rfloor }
\newcommand{\ovl}[1]{\overline{#1}}
\newcommand{\beq}{\begin{equation}}
\newcommand{\eeq}{\end{equation}}
\newcommand{\beqn}{\begin{eqnarray}}
\newcommand{\eeqn}{\end{eqnarray}}
\newcommand{\dd}{\mbox{d}}
\newcommand{\qb}{q}
\newcommand{\rr}{{\bf r}}
\newcommand{\xx}{{\bf x}}
\newcommand{\yy}{{\bf y}}
\newcommand{\ee}{{\bf e}}
\newcommand{\Sp}{\mathrm{Spec}}
\newcommand{\Tb}{\overline{T}}
\newcommand{\EE}{{\cal E}}
\newcommand{\HH}{{\cal H}}
\newcommand{\MM}{{\cal M}}
\newcommand{\FF}{{\cal F}}
\newcommand{\DD}{{\cal D}}
\newcommand{\RR}{{\cal R}}
\newcommand{\SS}{{\cal S}}
\newcommand{\nab}{\bbox \nabla}
\newcommand{\gam}{\bbox \gamma}
\newcommand{\tx}{\tilde{x}}
\newcommand{\epsun}{{\bf e}_1}

\begin{document}

\draft 
\title{Geometrical Approach for the Mean-Field Dynamics of a Particle
in a Short Range Correlated Random Potential}
\author{Fabrice{\sc Thalmann}$^{*,\dag}$} 
\address{LEPES-CNRS, Laboratoire associ\'e \`a l'UJF-Grenoble\\
{\it BP166X  25 avenue des Martyrs 38042 Grenoble Cedex} France\\
  and \\
$\dagger$ Department of Physics and Astronomy, University of Manchester\\
\textit{Oxford Road, Manchester M13 9PL} U.K.} 
 
\date{May~$4^{th}$, 2000}

\maketitle 

\begin{abstract}
  The zero temperature relaxational dynamics of a particle in a short
  range correlated random potential is addressed. We derive a set of
  ``two-times'' mean-field dynamical equations, accounting for a
  possible mean displacement of the particle when subject to an
  external force. We show first detailed results from the numerical
  integration of the above mentioned equations. We mainly pay
  attention to the exponentially decreasing spatial correlations case,
  for which simple analytical arguments provide valuable results about
  the hessian (or the ``instantaneous normal modes'' structure) of the
  energy landscape, and we propose a geometrical description of the
  ``mean-field aging''. Our numerical results and further analytical
  arguments give access to the waiting-time dependence of the main
  characteristic time scales.
\end{abstract}



\section{Introduction}
\label{sec:one}

Understanding of the out-of-equilibrium dynamics of glassy systems,
including spin glasses, structural glasses, supraconductor vortex
glasses, {\it etc}, is a challenging problem.  The need for exact, but
non trivial results led to the introduction of ideal spin-glasses,
like the celebrated Sherrington-Kirkpatrick model~\cite{SheKir}. The
spherical $p$-spin glass is such a model, where a closed set of
equations exists for the time correlation functions of the dynamical
variables, or soft spins~\cite{CriHorSom,CugKur}.  In the
thermodynamical limit, each spin becomes coupled to an infinity of
other spins. Each variable experiences gaussian fluctuations in the
effective environment created by all the other spins.  The dynamics
then simplifies drastically, and reduces to a set of correlation
functions which have to be determined self-consistently. The result
is the ``dynamical mean-field solution'' of the model.

The mean-field dynamics of spin-glasses has revealed extremely rich,
the most striking feature being the existence of a non-trivial aging
relaxation regime at low temperature~\cite{CugKur}. For instance, the
solution of the mean-field equations in this out-of-equilibrium regime
demonstrate the existence of a generalised fluctuation-dissipation
theorem (\textit{i.e.} connecting correlation and response functions)
whose validity seems now to extend to many realistic, non mean-field,
models~\cite{rem:FDT}.

Then, mean-field solutions are valuable for explaining the
experimental aging of disordered magnetic
systems~\cite{BouCugKurMez:2}. Finally, a sustained interest has
followed the discovery of a deep formal analogy between the mode
coupling description of structural glasses (supercooled liquids) and
the mean field treatment of the spherical $p$-spin
glass~\cite{KirThi}.

A crucial shortcoming of the mean-field description, however, is its
inability to take into account properly thermally activated motion
over energy barriers, leading to a sharp dynamical transition --
divergence of an internal relaxational time scale -- whereas the
corresponding ``finite dimensional'' behaviour is only a strong but
progressive slowing down of the dynamics.

Despite of this last point, mean-field dynamics remains a major issue
in the study of out-of-equilibrium statistical physics of disordered
systems, and any approach providing a physical insight on its aging
mechanism is of interest. A major step in that direction was made by
J.Kurchan and L.Laloux~\cite{KurLal} who investigated the zero
temperature relaxation of systems including ferromagnets and spin
glasses. The zero temperature limit makes it possible to consider the
energy landscape, rather than a ill-defined ``free-energy'' landscape,
without reducing the dynamics to anything trivial.

In this work, we extend further their approach, and apply it to
another system of interest: the mean-field dynamics of a particle in
a short-range correlated random potential.

The out-of-equilibrium, aging dynamics of this model has been first
studied in~\cite{FraMez:1,FraMez:2}, and thoroughly investigated
in~\cite{CugLeD}. Its glassy behaviour belongs to the same
universality class than the spherical $p$-spin model. What makes this
model interesting is its natural extension, when a finite and constant
external force is applied to the particle. Then, it becomes a paradigm
of ``driven glassy system'', in which a non-linear response to the
force as well as a significant violation of the
fluctuation-dissipation theorem are expected, as shown by
Horner~\cite{Horner}.

In this paper, we present the dynamical mean-field equations in the
presence of a constant force, allowing for an arbitrary mean
displacement along this one. These equations are then numerically
solved, in the zero temperature limit, for an exponentially decreasing
correlator, in the absence, and in the presence of a weak external
force. The corresponding numerical results are presented, restricting
ourselves to the linear response regime. Then, we start our
geometrical analysis of the zero temperature relaxation by a simple
random matrix calculation that we believe to describe satisfactorily
the hessian of the exponentially correlated gaussian potential. Next,
we perform an ``instantaneous normal mode'' analysis of the
relaxational motion. The key observable turns out to be the
(intensive) energy difference between the energy ${\mathcal {E}}(t)$
at a given time $t$, and its asymptotic value $\lim_{t \to \infty}
{\mathcal{E}} (t)$. We subsequently analyse the waiting time
dependence of two characteristic time scales $t_f,t_b$, that we relate
to $\EE(t)-\EE(\infty)$.

This work is preliminary to the study of the stationary driven
situation in the presence of a finite force, which will be the subject
of a forthcoming publication, and where the velocity-force
characteristics, and the cross-over between linear and non-linear
response will be exposed~\cite{ComingSoon}. 

   

\section{The out-of-equilibrium dynamics in the mean-field
approximation}
\label{sec:two}

We introduce in this section the mean-field dynamics equations and
discuss the low temperature aging solution in the absence of force.
Let ${\bf x}(t)$ be the position of a particle, obeying a usual
Langevin equation:
\begin{equation}
  \dot{{\bf x}}(t) = -{\nab} V({\bf x}(t)) + \FF
  +{\bbox \zeta}(t), \label{eq:Langevin}
\end{equation}
where are introduced the random potential $V({\bf x})$, the external
force ${\FF}$, a white Langevin noise ${\bbox \zeta}(t)$ corresponding
to a temperature $T$, and a friction coefficient equal
to~1. Quantities $\xx,\nab,\FF,{\bbox \zeta}$ are $N$-dimensional
vectors.  Three sub-cases of the dynamics defined
by~(\ref{eq:Langevin}) are of interest: 1/ the ``isolated'' dynamics,
without force: ${\FF}=0$; 2/ the driven relaxational dynamics, which
is the zero temperature limit of~(\ref{eq:Langevin}): ${\FF} \neq 0$
and $T \to 0$; 3/ the relaxational ``isolated'' dynamics: ${\FF} =0,
T\to 0$.

The potential $V({\bf x})$ is a quenched disorder, chosen from a
gaussian distribution.  All the averages with respect to it will be
denoted with an over-line~$\ovl{\hspace{0.2cm}\cdot\hspace{0.2cm}}$,
while the average over the thermal noise (if any) ${\zeta}$ will be
denoted by the brackets~$\mean{\cdot}$.  We suppose that the motion
starts at $t\!=\!0$ and ${\bf x}(t\!=\!0)\!=\!0$.  After averaging
over the quenched disorder, this choice becomes equivalent to start
with a random, ``infinite temperature'' distribution of initial
positions. We expect that the process~(\ref{eq:Langevin}) is
self-averaging with respect to $V({\bf x})$ in the infinite
dimensional limit.  One introduces the correlator $f(y)$ of the
gaussian disorder, explicitly dependent on the dimension $N$ of the
configuration space~$\{ {\bf x} \}$.
\begin{equation}
  \ovl{V({\bf x})\cdot V({\bf x}')} = N\cdot
  f\left(\frac{\|{\bf x}-{\bf x}'\|^2}{N} \right)\ ;\ 
  \ovl{V({\bf x})} = 0. \label{eq:correlator}
\end{equation}
This form ensures a meaningful $N\to \infty$ limit, in which each
coordinate $\xx_i(t)$, or gradient component $\partial_i V({\bf x})$,
remains of order~1, while the norms $\|{\bf x}(t)\|$, $\|{\nab} V\|$
scale like ~$N^{1/2}$. As a consequence, the external force must scale
($\epsun$ being a unit vector) like:
\begin{equation}
  {\FF} = N^{1/2}\cdot F \cdot \epsun.
  \label{eq:external-force}
\end{equation}
One expects a displacement $\ovl{\mean{{\bf x}(t)}}= N^{1/2}\cdot
u(t)\cdot \epsun$, and possibly a mean velocity $\ovl{\mean{\dd{\bf
x}(t)/\dd t}}=N^{1/2}\cdot v\cdot \epsun$. From now onwards, we
arrange that $\epsun$ coincides with the first coordinate axis $i=1$.

In the present paper, we restrict ourselves to the exponentially
correlated potential:
\begin{equation}
  f(y) = \exp(-y). \label{eq:exponential:correlator}
\end{equation}

This is a special case of short range correlated random potential,
characterised by $\lim_{y\to \infty} f(y) < \infty$. The average
difference $\ovl{[V({\bf x})-V({\bf x}')]^2}$ is bounded when $\|{\bf
x}-{\bf x}'\|$ grows, and this ensures the existence of a normal
diffusion regime at temperatures high enough. Another common choice is
the power-law correlator: 
$ f(y) = 2/(\gamma-1)\cdot(1+y)^{(1-\gamma)/2};\ \gamma >1$
\cite{FraMez:1,FraMez:2,CugLeD,Horner}. 
Choice~(\ref{eq:exponential:correlator}) is also a particular case of
$f(y) = U_p^2 \exp(-y/\xi^2)$, with a pinning energy $U_p$ and
correlation length $\xi$ set to 1, thanks to a simple rescaling,
without loss of generality.

The Langevin dynamics is handled with the help of a Martin-Siggia-Rose
(MSR)-like functional integral, convenient for averaging over the
gaussian disorder~\cite{DeDominicis,MarSigRos}. All the technical
details corresponding to the saddle-point equations as $N\to \infty$,
are given in~\cite{KinHor:1} and the action is given in
appendix~\ref{app:one}.  The crucial point is that the limit $N\to
\infty$ is taken first, before any other limit $T\to 0$ or $t \to
\infty$.  As a result, we obtain a general effective quadratic action
$S[x_j(t),i\tx_j(t)]$, involving the original field $x_j(t)$, and the
MSR auxiliary field $i\tilde{x}_j(t)$. Three among the four following
correlation functions appear explicitly in the action
$S[x_j(t),i\tx_j(t)]$:
\begin{eqnarray}
  u(t) & = & N^{-1/2}\ \ovl{\mean{ x_1(t) }}; \\
  r(t,t') & = & N^{-1}\sum_{j=1}^{N} \ovl{\mean{ x_j(t)\cdot
  i\tx_j(t') }}; \\
  b(t,t') & = & N^{-1}\sum_{j=2}^{N} \ovl{\mean{ (x_j(t)-x_j(t') )^2}}; \\
  d(t,t') & = & N^{-1}\sum_{j=1}^{N} \ovl{\mean{ (x_j(t)-x_j(t') )^2}};
  \nonumber\\ 
          & = & b(t,t') + [u(t)-u(t')]^2. \label{eq:motion:equations}
\end{eqnarray}
These are the displacement $u(t)$, the response function $r(t,t')$,
and the correlation functions $b(t,t')$ and $d(t,t')$.  The Dyson
equations for $r,b,d,u$ form a closed system of coupled
integro-differential equations. For $t\ge t'$ one has to solve:

\begin{eqnarray}
  \partial_t r(t,t') & = & \delta(t-t') \nonumber\\
  & & \hspace{-1cm} -\int_0^{t} \!\dd s  \ 4f''(d(t,s))\ r(t,s)\
  [r(t,t') - r(s,t')];     \label{eq:Dyson-2temps:r}\\
   \partial_t b(t,t') & = & 
    2T -\int_0^{t} \! \dd s \ 4f'(d(t,s)) \ [r(t,s) - r(t',s)]
   \nonumber\\ 
                      &   &  \hspace{-2cm}
   - \int_0^{t} \! \dd s \ 4f''(d(t,s)) \
   r(t,s) \ [b(t,s) + b(t,t') -b(s,t')];  \label{eq:Dyson-2temps:b}\\
   \partial_t u(t)    & = & F - \int_0^{t} \!\dd s \ 4f''(d(t,s)) \
   r(t,s) \cdot [u(t) - u(s)].\nonumber\\
   &&  \label{eq:Dyson-2temps:u} 
\end{eqnarray}
Equations~(\ref{eq:Dyson-2temps:r}-\ref{eq:Dyson-2temps:u}) are
original ones, and allow for a non uniform displacement $u(t)$. The
aging, isolated, situation corresponds to the limit
$u(t)\!=\!F\!=\!0$, and $d(t,t')\equiv b(t,t')$ in the above system.
The stationary limit, investigated by Horner~\cite{Horner} amounts to
write $r(t,t')=R(t-t')$, $b(t,t')=B(|t-t'|)$, $u(t)=v\cdot t$, and
to reject the lower bound of the time integrals $\int\dd t\ \dd s$ to
$-\infty$.

Three relevant observables: the energy ${\mathcal{E}}(t)$, the
curvature ${\mathcal{M}}(t)$ and the pinning force
$F_p(t)$ can be expressed with the help of these correlation
functions.
\begin{eqnarray}
  \EE(t) & = & N^{-1}\ \ovl{\mean{V({\bf x}(t))}}, \nonumber \\
         & = & \int_{0}^{t}\dd s\ 2f'(d(t,s))\ r(t,s); 
         \label{eq:dynamical-energy}\\
  \MM(t) & = & N^{-1} \sum_{j=1}^{N} \ovl{\mean{ \partial^2_{jj}
  V({\bf x}(t)) }}, \nonumber \\
         & = & \int_{0}^{t}\dd s\ 4f''(d(t,s))\ r(t,s); 
         \label{eq:dynamical-curvature}\\
  F_p(t) & = & N^{-1/2}\ \ovl{\mean{ -\partial_1 V({\bf x}(t)) }},
  \nonumber\\
         & = & -\int_{0}^{t}\dd s\ 4f''(d(t,s))\ r(t,s)\ [u(t)-u(s)]. 
         \label{eq:dynamical-friction}
\end{eqnarray}
The pinning force is such that $\ovl{\mean{\dd{\bf x}(t)/\dd t}} =
{\FF}_p(t)+{\FF}$ with ${\FF}_p(t) = N^{1/2}\cdot F_p(t)\cdot \epsun$.
The pinning force $F_p$ and the driving force $F$ have opposite signs.

A proper study of the mean-field equilibrium phase diagram requires an
extra quadratic confinement potential $\mu{\bf x}^{\,2}/2$.  This
ensures the existence of a true thermal equilibrium in the high
temperature phase, while correlation functions reach their asymptotic
values exponentially fast.  Then, a transition line $T_d(\mu)$, called
dynamical temperature, separates the high temperature ergodic phase,
from a low temperature, aging and non ergodic
phase~\cite{KinHor:1,FraMez:1,FraMez:2}.

At high temperature, the system reaches a true stationary state, and
the dynamics becomes time-translationally invariant (TTI), {\it i.e}
the 2-times correlation functions depend only on the difference
$t-t'$, while the 1-time expectation values are constant.  In this
stationary situation, it is convenient to introduce the TTI
correlation functions $B(t-t')=\lim_{t,t' \to \infty} b(t,t')|_{t-t'\
\mathrm{finite}};\ R(t-t')= \lim_{t,t' \to \infty} r(t,t')|_{t-t'\
\mathrm{finite}}$. The fluctuation-dissipation theorem (FDT) holds and
reads:
\begin{equation}
  \dd B(t)/ \dd t =  2T\cdot  R(t).
\end{equation}
As a consequence, the equal-time correlation functions coincide with
their thermodynamical (canonical ensemble) counterparts.  Taking the
limit $\mu \to 0$ does not lead to any singular
result~\cite{CugLeD}. Provided the contribution from the harmonic
potential has been subtracted off, the energy $\EE(t)$ behaves
smoothly as $\mu$ tends to~0.  When $\mu$ exactly equals~0, the system
cannot be at equilibrium, and instead, one expects a long time
behaviour corresponding to a normal diffusion situation, with a finite
diffusivity $D = \lim_{t \to \infty} \mean{{\bf x}^{\, 2}(t)} /(2Nt)$,
a finite mobility $\eta^{-1} = \lim_{t \to \infty} u(t)/(Ft)$, and the
Einstein relation $D=T\cdot \eta^{-1}$. 

Kinzelbach and Horner described the dynamics in the stationary, high
temperature phase~\cite{KinHor:1}. They found that these correlation
functions behave in the same way than those of the well known
mode-coupling theories for supercooled liquids, as expected on general
grounds~\cite{rem:MCT,BouCugKurMez}.\nocite{Gotze} The non-linearities
of the self-consistent equations cause a dramatic slowing down of the
dynamics as $T_d$ is approached from above, leading to a sharp
transition at $T=T_d$.

For instance, the function $B(t)$, after a fast increase at short
times $t\sim 1$, has a long plateau near a characteristic value
$B(t\sim t_f) \simeq q$, before eventually reaching its asymptotic,
long time regime $B(t) = \hat{B}(t/t_b)$. Both $t_f$ and $t_b$ diverge
like power laws of the difference $|T-T_d|$~\cite{KinHor:1}.

The low temperature region however corresponds to an
out-of-equilibrium situation.  In the absence of external force, this
is meant by the loss of both time-translational invariance (TTI)
and fluctuation-dissipation theorem (FDT). The 2-time correlation
functions cannot be reduced any more to functions of the time
differences $t-t'$, and there is a domain in the $(t,t')$ plane, where
the system ages~\cite{FraMez:1,FraMez:2,CugLeD}.

The addition of a weak, constant external force leads to a somewhat
different picture. As proposed by Horner, the system is expected to
reach a stationary state (TTI), but the FDT remains definitively
lost~\cite{Horner}. It turns out, however (cf next section) , that
when the force is switched on at a time $t=0$, there is a finite time
interval during which the dynamics can be successfully described as a
perturbation around the aging isolated ($F=0$) regime, with a linear
response approach.  The extent of this linear response regime is
inversely related to the magnitude of the force.

The aging dynamics of the isolated particle has been exhaustively
treated in~\cite{CugLeD}. The fluctuation-dissipation theorem is
violated and must be replaced by:
\begin{equation}
\label{eq:qfdt:1}
  X(t,t') \partial_{t'} b(t,t') = r(t,t').
\end{equation}
In the time sector $t-t'$ finite; $t' \to \infty$ of the $(t,t')$
plane, the behaviour is very similar to the one observed just above
$T_d$, and the value of $X(t,t')$ is very close to its equilibrium
value $-1/(2T)$. When the time separation $t-t'$ ceases to be small
relatively to a characteristic time $t_f(t')$ which has to be
determined, $X(t,t')$ departs from its equilibrium value, decreasing
its magnitude $|X|$.

The analytical study of the equations~(\ref{eq:motion:equations}) has
only been possible in the asymptotic limit $t,t' \to \infty$, by
dropping out sub-leading terms presumably of order $1/t,1/t'$. In this
limit, the authors have shown (this is the crucial point) that it was
possible to parametrise the dynamics with the help of the correlation
function $b(t,t')$ of the system itself. This implies that $X(t,t')$
becomes a one variable function $X[b(t,t')]$, and it turns out that
all short range correlated models can be solved thanks to the
ansatz~\cite{CugLeD,CugKur:2}:
\begin{eqnarray}
\label{eq:qfdt:2}
  b(t,t') < q\ & \Rightarrow & X[b] = -1/2T; \nonumber\\ b(t,t') > q\
  & \Rightarrow & X[b] = \chi.
\end{eqnarray}
This extension of the FDT is called ``quasi-fluctuation-dissipation
theorem'' (QFDT).  In this paper, we rather use the function
$\Tb(t,t')=-1/(2X(t,t'))$.  In the aging regime, this effective
temperature $\Tb=-1/(2\chi)$ is higher than the thermostat temperature
$T$, and remains finite in the limit $T\to 0$. The physical
meaning of these ``two or many temperatures systems'' is discussed
in~\cite{CugKurPel}.

In the case we are interested in, \textit{i.e.} in the absence of
confinement ($\mu \to 0$), for a correlator $f(y)=\exp(-y)$, $\chi$
and $q$ are for any temperature $T<T_d$, solutions of the
system~\cite{CugLeD}:
\begin{equation}
\label{eq:value-qx}
\left\lbrace
\begin{array}{ccccc}
T & = & q\sqrt{f''(q)} & = & q\ e^{-q/2}; \\
\chi & = & \frac{\sqrt{f''(q)}}{2f'(q)} & = & -e^{q/2}/2;
\end{array}
\right.
\end{equation}
and in the low temperature limit:
\begin{equation}
\label{eq:value-qx-lowT}
\left\lbrace
\begin{array}{ccc}
q & \simeq & T+T^2/2 \ldots; \\
\chi & \simeq & -\left( \frac{1}{2} + \frac{T}{4} +\frac{3T^2}{16}\ldots
\right) 
\end{array}
\right.
\end{equation}
In the same way, given a triplet $t_1<t_2<t_3$, in the time domain
where $(t_1,t_2,t_3) \to \infty$ and $t_1/t_2$, $t_2/t_3$ finite,
$b(t_3,t_2)$ is uniquely determined by the knowledge of $b(t_2,t_1)$
and $b(t_3,t_1)$. Again, the explicit dependence can be carried out
exactly when the correlator is exponential. The result
is~\cite{CugLeD}:
\begin{equation}
  b(t_3,t_2)-q = b(t_3,t_1)-q - [b(t_2,t_1)-q].
\end{equation}
A well known shortcoming of this approach, is that any reference to
the original times $t,t'$ is definitively lost. The asymptotic
solution cannot distinguish between $b(t,t')$ and $b(h(t),h(t'))$
where $t \mapsto h(t)$ can be any suitable reparametrization of the
time variable.  As a by-product, the previous analysis predicts only
the more general form of the solution, in the aging regime $t/t'\sim~1$:
\begin{equation}
   b(t,t') = \tilde{B}\left[\ln\left( \frac{h(t)}{h(t')} \right)\right] + q.
   \label{eq:scal-form-ageing}
\end{equation}
For exponentially correlated potentials, the master function is
known~\cite{CugLeD}, and without loss of generality:
\begin{equation}
   b(t,t') = \ln(h(t))-\ln(h(t')) + q.
   \label{eq:special-form-b} 
\end{equation}
In~\cite{CugLeD} is made the conjecture $h(t)= t^{\delta}$, compatible
with the results found below. In what follows, we will refer to this
solution as the time-reparametrization invariant (TRI) solution.

At the beginning of the aging regime, for $t$ and $t'$ such that
$(t-t')/t'$ is finite but small compared to~$1$, the scaling
form~(\ref{eq:scal-form-ageing}), reads:
\begin{equation}
     b(t,t') = \tilde{B}\Big((t-t')\cdot t_b^{-1}(t') +\ldots
     \Big) + q.    
\label{eq:definition-tb}
\end{equation}
Here, $t_b(t')$ is the characteristic time of the aging regime,
defined by $t_b(t') = h(t')/h'(t')$. This is the typical time needed
by the particle for diffusing over a distance $b(t,t')-q \sim 1$. Non
exponential correlators have a non analytic scaling function
$\tilde{B}(m)$ around $m=0$ and the r.h.s
of~(\ref{eq:definition-tb}) is singular in
$t-t'$~\cite{CugLeD,CugKur:2}.

The time-reparametrization invariant solution describes a situation
where the time scales for the FDT regime ($t_0=1$) and for the aging
regime ($t_b(t')$) are well separated ({\it i.e.} $t_b(t') \gg 1$),
which implicitly assumes $t,t' \to \infty$.  In order to go further,
one needs to take into account the times derivatives
$\partial_t,\partial_{t'}$ neglected in the asymptotic regime of the
TRI solution.  It is enough, in principle, to fix up the
reparametrization function $h(t)$.  Moreover, the TRI solution does
not say how the parameter $\Tb$ goes from its FDT value ($b(t,t')<q$)
to its QFDT value ($b(t,t') > q$). One defines for this purpose the
new time scale $t_f(t)$, such that, for instance, $\Tb(t,t-t_f(t))$
takes a given value between $T$ and $-1/(2\chi)$.  We shall see below
that $t_f(t)$ is much smaller than $t_b(t)$.


\section{Results from the numerical integration of mean-field 
equations}
\label{sec:three}

The mean-field equations, with $F$ and $u(t)$ equal to zero were first
numerically integrated by Franz and
M{\'e}zard~\cite{FraMez:1,FraMez:2}. The quadrature scheme is of order
one in the time step $h$, but reveals itself surprisingly robust as
$h$ is increased up to value as large as $0.3$. The authors
of~\cite{FraMez:1} report being able to reach $t\sim 1000$ at the
best. Our investigations have shown that the quality of our solutions
gets worst if $h$ in increased above $0.2$, and we present results up
to $t\sim 400$.

For reasons detailed in the next section, we have only considered the
exponential correlator case~(\ref{eq:exponential:correlator}). We set
$T$ to 0 in~(\ref{eq:motion:equations}) and took the initial value
$C(0,0)=0$.  The information coming from the numerics may be
pigeonholed in three categories.

\begin{figure}
    \epsfxsize 12cm  \centerline{
    \epsfbox{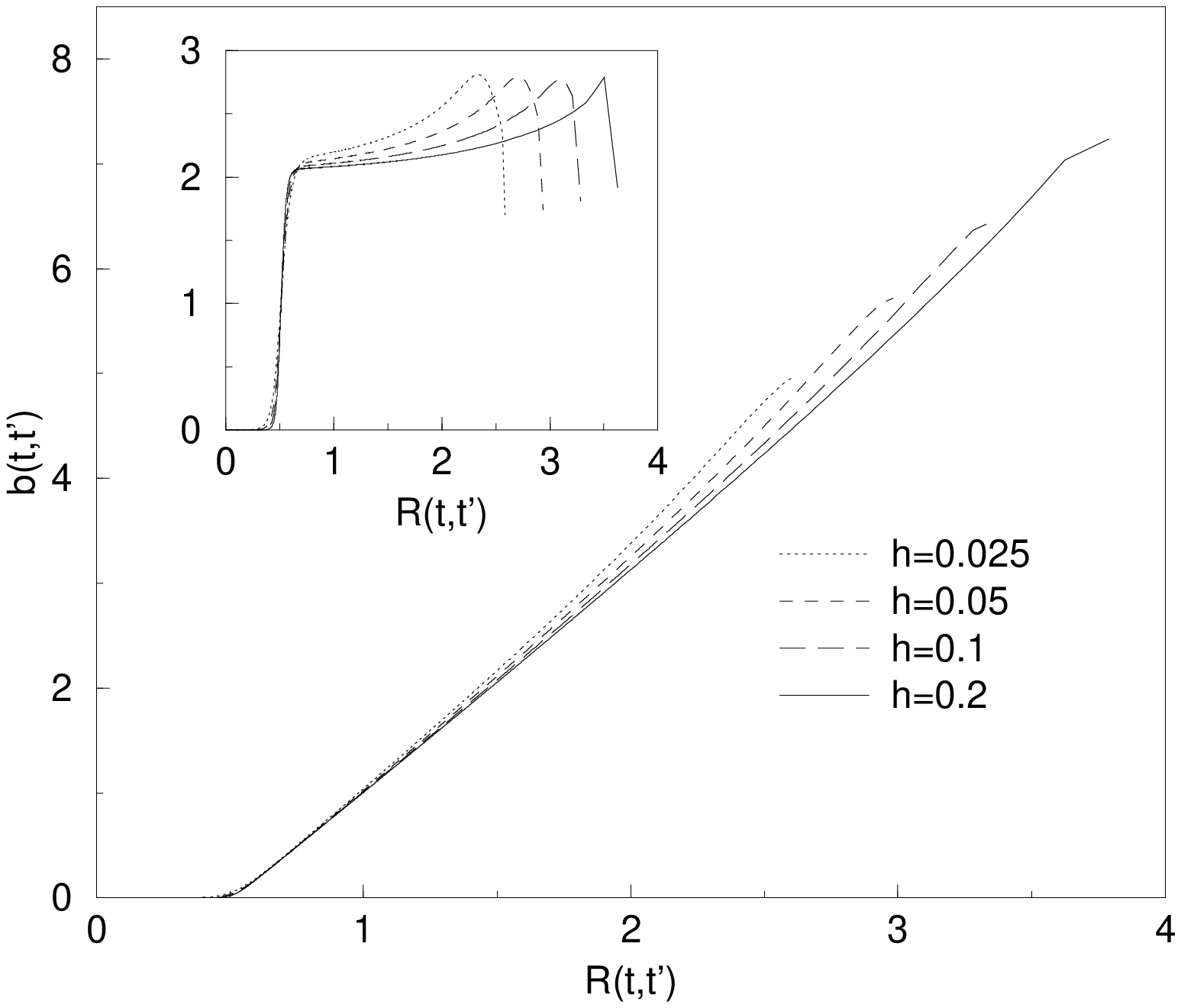} } 
    \caption{ }
    \label{fig:x-of-b}
\end{figure}

\textbf{1/ Results related to the TRI solution.} First of all, we must
check that the quasi-fluctuation dissipation
relation~(\ref{eq:qfdt:2}) is true by plotting the integrated response
versus the correlation function, on Figure~(\ref{fig:x-of-b}). The
observed value of $\chi$ is close to 0.46, while the predicted value
is $1/2$. The TRI solution predicts also $q\simeq 0$ and $\lim_{t\to
\infty} b(t,0)=\infty$, in the absence of confinement. The measured
asymptotic energy ${\mathcal{E}}(\infty)$ and mean curvature
${\mathcal{M}}(\infty)$ are found to be in excellent agreement with
the predicted values $-2$ and $+4$ respectively.

\begin{figure}
    \epsfxsize 12cm  \centerline{
    \epsfbox{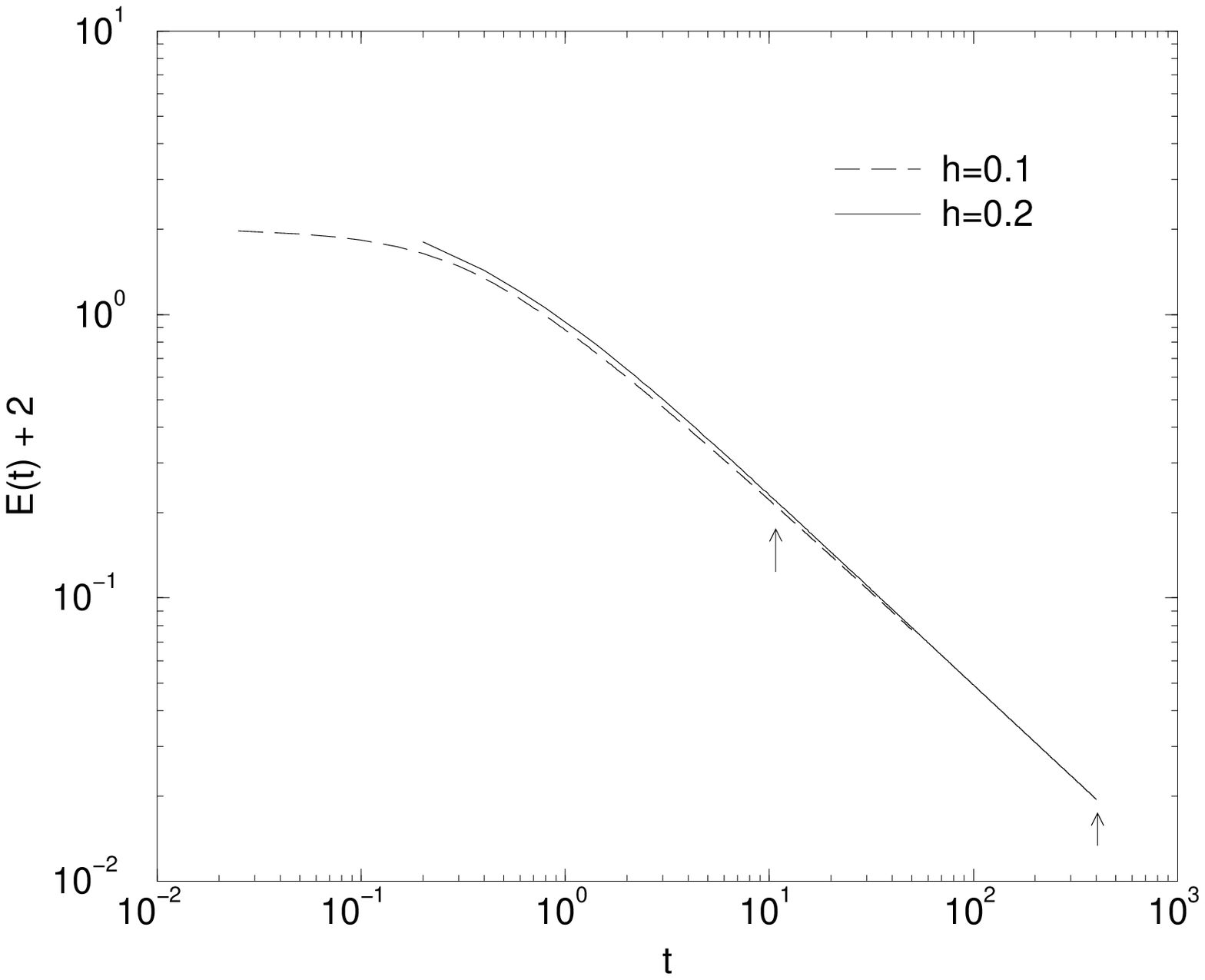} } 
    \caption{ }
    \label{fig:energy}
\end{figure}

\textbf{2/ Beyond the TRI solution, without external force.} This includes
for instance the algebraic decay of the energy ${\EE}(t)= -2 +
c_1\cdot t^{-\kappa}$. The exponent $\kappa$ is determined by plotting
$\log(2 + {\EE(t)}) $ versus $\log(t)$, and also by computing directly
the logarithmic derivative, as shown on Figure~(\ref{fig:energy}). The
exponent $\kappa$ lies between $0.66$ and $0.67$ and our best estimate
is $c_1=1.08$.

Also concerned are the characteristic times of the aging regime, and
the precise nature of the cross-over from equilibrium to
quasi-equilibrium fluctuation dissipation theorem. We are interested
here in finding the characteristic time $t_f(t)$ as a function of $t$,
defined by:
\begin{equation}
\label{eq:tf}
  \int_{t-t_f}^{t} \!\dd s\ 2 f'(b(t,s))\ r(t,s) = -1,
\end{equation}
or alternatively:
\begin{equation} 
\label{eq:tf-prime}
  \int_{t-t_f'}^t \! \dd s\  r(t,s) = -\frac{1}{2f'(0)}= \frac{1}{2}.
\end{equation}
Equation~(\ref{eq:tf}) comes from the fact that the equilibrium
$X=-1/2T$ and aging $X=\chi$ time sectors contribute for $-1$ each to
the energy. The value $t_f$ which solves~(\ref{eq:tf}) separates the
equilibrium regime ($b(t,s) < b(t,t-t_f)$) from the aging one ($b(t,s) >
b(t,t-t_f)$). The equivalence between~(\ref{eq:tf}) and
~(\ref{eq:tf-prime}) is a straightforward consequence
of~(\ref{eq:qfdt:2})

One generalises~(\ref{eq:tf-prime})~in:
\begin{equation}
\label{eq:ta}
  \int_{t-t_a}^{t} \!\dd s\  r(t,s) = a,
\end{equation}
For $a<1/2$, $t_a$ must tend to a constant as $t \to \infty$, while
for $a>1/2$, the asymptotic scaling~(\ref{eq:scal-form-ageing})
predicts:
\begin{eqnarray}
\label{eq:ta:2}
  a-1/2 & = &  -\chi\ \tilde{B}\left[\ln\left(
  \frac{h(t)}{h(t-t_a)}\right)\right]; \nonumber \\
  & \simeq & -\chi\  \tilde{B}(t_a/t_b),
\end{eqnarray}  
where terms $(t_a/t_b)^2$ have been neglected in the last expression, and
$t_b=h(t)/h'(t)$. If $a$ is small enough, $t_a$ is simply proportional
to $t_b$.  Moreover, if $h(t)$ is indeed $t^{\delta}$, then $a-1/2=
-\chi\ \tilde{B}_1( (1-t_a/t_b)^{-\delta})$, and $t_a/t_b$ is strictly
constant.  Our Figure~(\ref{fig:tf}) shows a plot of $t_f$,
$t_{a=0.55}$, and $t_{a=0.45}$. The characteristic time scale $t_f$
tends asymptotically towards a power law $c_2 t^{\alpha}$, with $c_2
\simeq 0.51$ and $\alpha\simeq 0.64$ (according to our best
estimate).

The correlation function is found to grow logarithmically with $t$,
and $f(b(t,t'))= \exp(-b(t,t'))$ behaves as a power law of
$t$. Figure~(\ref{fig:exp-b}) presents $\exp(-b(t,0))$ and
$\exp(-b(t,t'))$ for a fixed $t'$. An algebraic decay $t^{-\delta}$
of $\exp(-b(t,0))$ is likely, while $\exp(-b(t,t'))$ has not yet
reached its asymptotic regime, but could tend to the same
$t^{-\delta}$ behaviour.

From the asymptotic form~(\ref{eq:scal-form-ageing},
\ref{eq:special-form-b}) we note that $\exp(-b(t,t')) \simeq
h(t')/h(t);\ t,t' \to \infty;\ t/t'$ finite, and our Figure is
consistent with $h(t) = t^{\delta}$, $\delta \simeq 1.10$.  Also
shown is $\exp(-b(t,t-t_f)) = t^{-\gamma}$, $\gamma\simeq
0.42$. Expanding $\exp(-b(t,t_f))$ as $(t/t_f)^{-\delta}$, and using
$t_f \sim t^{\alpha}$, one finds a relation $\gamma=\delta(1-\alpha)$
between exponents. The agreement between $\delta(1-\alpha)=0.39$ and
$\gamma\simeq 0.42$ is acceptable.

\textbf{3/ The linear displacement regime, in the presence of a
driving force.}  
A small force $F$ is applied and the displacement $u(t)$
monitored. The linear response implies that $u(t)$ must be
proportional to $F$, and it is indeed the case for time intervals not
too large. Figure~(\ref{fig:log-disp}) presents $u(t)/F$ for
decreasing values of $F$. The curve $F=0.05$ is virtually
indistinguishable from the integrated response $\RR(t) = \int_0^t
r(t,s) \dd s$, and this shows that $\lim_{F \to 0} u_F(t)/F = \RR(t)$,
\textit{i.e.} the expected linear response behaviour. The other curves
depart from the integrated response after a time $t_F$ decreasing with
$F$. When starting from the isolated and aging situation $F=0$, the
linear response only holds during a finite time interval $0< t \le
t_F$. What happens later is the onset of a stationary state, with a
well defined velocity $v$ and a non-linear dependence in the force as
advocated by Horner~\cite{Horner}. A study of this regime is to be
published soon~\cite{ComingSoon}.

In the linear response regime, Figure~(\ref{fig:log-disp}) is compatible
with:
\begin{equation}
\label{eq:u:linear-response}
  u(t) = F \cdot ( c_3 + c_4 \cdot \ln(t) ), 
\end{equation}
$c_3\simeq 0.71$ and $c_4=0.54$.


\section{The geometrical approach}
\label{sec:four}

In this section, we transpose to the particle in a random potential
some of the ideas which have revealed fruitful when applied to the
spherical $p$-spin model, namely the geometrical analysis of Kurchan
and Laloux~\cite{KurLal}. We expose first the main concepts of the
method, and then propose a method for computing the limit value of the
dynamical energy from generic properties of the potential~$V({\bf
x})$, working only for an exponential correlator~$f(y)=\exp(-y)$.

The $p$-spin model starts aging below a dynamical temperature
$T_d$~\cite{CriHorSom}, and encounters a thermodynamical glassy
transition at $T_s < T_d$~\cite{CriSom}. Detailed investigations have
brought an appealing picture of the complex free-energy landscape of
the spherical $p$-spin model, accounting for many features of its
thermodynamics~\cite{KurParVir,FraPar,CavGiaPar:1} and its
dynamics~\cite{CugKur,BarBurMez:2}.

The phase space of the $p$-spin model can be investigated with the
help of a ``Thouless-Anderson-Palmer'' free-energy $\Phi(m_i)$ of the
magnetization $m_i, i=1\ldots N$. At low enough temperatures, the
function $\Phi$ develops many extrema $m_i^{(\alpha)}$, the TAP
solutions $\alpha$. Those extrema which are minima, {\it i.e.} the
second derivative matrix $\partial^{2}\Phi/\partial m_i\partial m_j$ is
definite positive, are metastable states, as they are separated from
each others by extensive free-energy barriers. A particular
realization of the system, prepared in a given metastable state
$\alpha$ remains for ever in this state in the thermodynamic limit.

The stability of a metastable state is related to the lowest
eigenvalue $\lambda_{min}$ of $\Sp(\partial^{2}\Phi/\partial
m_i\partial m_j)$, spectrum of the hessian matrix. $\lambda_{min}$
turns out to be a monotonically decreasing function of the free energy
$\Phi(m_i^{(\alpha)})$ of the state itself. This defines the
free-energy $\Phi_d$ of the marginal states as $\lambda_{min}=0$ for
$\Phi=\Phi_d$. Magnetizations such that $\Phi(m_i) \geq \Phi_d$
represent regions of negative curvature which does not contribute to
the thermodynamics but play a role in the dynamics~\cite{Biroli}.

The glassy dynamics of the $p$-spin model is observed when the stable
metastable states $\Phi^{(\alpha)} < \Phi_d$ are populated, or
equivalently, when the canonical Boltzmann measure is split into its
metastable components $\alpha$. At a temperature lower than $T_d$,
thermal equilibration requires the system to explore all the relevant
metastable states $\Phi<\Phi_d$. Such an equilibration is impossible
as going from one metastable state to the next one requires to go over
an infinite barrier. What happens instead to a system quenched from
high temperature, to $T<T_d$, is the onset of aging. The systems wander
more and more slowly around the magnetization region $\Phi(m_i) \simeq
\Phi_d$, {\it i.e.} around the marginal states. 

The zero temperature relaxational dynamics is simpler because the free
energy reduces to the hamiltonian $H$ of the spins $s_i$. At variance
with the finite temperature case, the regions with negative curvature
of $H$ are now well defined.  Taking the limit $T\to 0$ in the
mean-field equations does not lead to any singular behaviour. This
somewhat counter-intuitive property is the consequence of sending
$N\to \infty $ first, by keeping finite the times $t$ and $t'$.  The
dynamics is a pure gradient descent, but remains non trivial.

 In order to perform a geometrical analysis of this relaxational
dynamics, it is necessary to keep the dimension~$N$ large but finite.
Then, the relaxational process occurs till the particle falls into a
true minimum of the Hamiltonian $H(s_i)$, and, at $T=0$, remains stuck
there indefinitely. According to the description advocated
in~\cite{KurLal}, a system starting from a random
configuration~$\{s_i(0)\}$ will explore regions with smaller and
smaller gradient $\|{\nab} H(s_i)\|$, and a decreasing number of
negative eigenvalues in the spectrum of~$\HH_{ij}$, hessian of
$H(s_i)$. The typical time $t_N^I$ for reaching regions where
$\HH_{ij}$ has only $I$ negative eigenvalues, diverges as $N$ is sent
to $\infty$ by keeping $I$ finite~\cite{KurLal}.  As a consequence, in
the $N\to \infty$ limit, the system is unable to reach within a finite
time $t$ a true minima, or even a saddle between two minima, and the
difference $\EE-\EE_d$ remains positive.

To what extent does the above picture describe the particle in a
random potential~?  Much less is known about the properties of the
metastable states, and there is no obvious equivalent of the free
energy functional $\Phi (\mean{{\bf x}})$ of the mean particle's
position $\mean{{\bf x}}^{(\alpha)}$ in the state
$\alpha$. Nonetheless, we expect that the basic mechanism of the
dynamical transition remain the same as for the spherical $p$-spin
model, {\it i.e.} a slow relaxation toward a region of marginal
states, $\lambda_{min} \sim 0$.  When considering the zero temperature
limit, the dynamics reduces to a gradient descent $\dot{\xx}(t) 
= -{\nab}V({\bf x}(t))$. The metastable states now correspond
to local minima of the potential $V({\xx})$, and their stability
will depend on the spectrum of the hessian $\HH_{ij} = \partial^2_{ij}
V({\bf x})$, where $\partial_i$ means $\partial /\partial x_i$ and
$\partial^2_{ij} = \partial^2/\partial x_i \partial x_j$.

The purpose of the geometrical approach, at zero temperature, is to
relate the values provided by the more formal field-theoretical
approach, to basic properties of the potential $V({\bf x})$.  For
instance, one must be able to compute the asymptotic values of the
energy $\EE(t)$, curvature $\MM(t)$, and the ``plateau value''
$\lim_{T\to 0}\qb/T$ of the correlation function. This is the first
step, already outlined in~\cite{KurLal}. One of the original
contributions of this work concerns a second step, where we justify,
with some geometrical arguments, many of the fine properties of the
aging behaviour, beyond the time reparametrization invariant solution.

\textbf{A random matrix computation of the spectrum of $\HH$.}
A challenging problem in the study of the supercooled liquids
dynamics, concerns the computation of the canonically averaged
spectrum of the instantaneous normal modes
$\mean{\Sp(\HH)}$~\cite{Keyes}. Here, $\HH$ is the dynamical matrix,
hessian of the potential energy $V$ of the interacting particles,
function of the coordinates $\mathbf{r}_i$. $\Sp(\HH)$ is any
representative characteristic function of the eigenvalues spectrum,
\textit{e.g.} the density of states.
\begin{equation}
\label{eq:inm-definition}
  \mean{\Sp(\HH)} = Z^{-1}\cdot\int_{\mathcal{D}}\dd \xx\ \left( e^{-\beta
  V}\cdot \Sp(\HH) \right).
\end{equation}

$Z$ is the canonical partition function, and $\DD$ is a bound domain
eventually becoming infinite. In the mean field situation, as the
energy $V$ is extensive, the canonical average is dominated by a
saddle point value $V(\beta)$ of the potential $V(\xx)$. The analogous
of (\ref{eq:inm-definition}) becomes:
\begin{eqnarray}
	\mean{\Sp(\HH)} & = & W_{\beta}^{-1}\int_{\DD} \dd {\mathbf x}\
	\delta (V({\mathbf x})-V(\beta))\ \Sp(\HH)({\mathbf x}),
	\nonumber\\ 
	W_{\beta} & = & \int_{\DD} \dd{\xx}\ \delta (V(\xx)-V(\beta)). 
\label{eq:slice-nm-definition}
\end{eqnarray}

In the $p$-spin case, the spectrum of $\HH$ is self-averaging with
$\xx$, \textit{i.e.} the spectrum of $\HH$ is a shifted
semi-circle, by an amount controlled by $V(\beta)$ only.  The
dynamical energy is found to be the highest value of $V(\beta)$ such
that $\Sp(\HH) \geq 0$ (marginality).

The averaged spectrum $\mean{\Sp(\HH)}$ defined by
equation~(\ref{eq:slice-nm-definition}) seems to be the natural
quantity to consider when looking at the zero-temperature relaxational
dynamics of our mean field model. We have found that as far as
exponentially correlated potentials are concerned , $\Sp(\HH)$ is, at
the leading order, a non-fluctuating quantity determined by
$V_0=V(\beta)$.  More precisely, $\Sp(\HH)$ has got a semi-circular
distribution of radius $\Lambda$, centred around $\DD$.
\begin{eqnarray}
       \Lambda & = & 4 \sqrt{f''(0)};
\label{eq:width:hessian} \\
       \DD & = & \frac{2f'(0)}{f(0)} \frac{V_0}{N}.
\label{eq:shift:hessian}
\end{eqnarray}

Let us outline our demonstration. We consider first the (${\rr}$
independent) ``annealed average''.
\begin{equation}
\label{eq:1point:annealed}
   \ovl{ \delta(V(\rr)-V_0)\ \Sp(\HH)(\rr) }.
\end{equation}
In order to compute (\ref{eq:1point:annealed}), it is enough to
enumerate the correlations of $V(\rr), \partial_{ij}V(\rr)$, where
$\rr$ is an arbitrary point. All the $\partial_{ij}V(\rr)$ are
independent at the leading order $N^{-1}$, whereas the $N\!+\!1$
remaining variables $V(\rr), \partial_{ii}V(\rr)$ are found to be
correlated. One has:
\begin{equation}
\label{eq:correlations:V}
\begin{array}{cccc}
  N\cdot \ovl{ [\partial_{ij}V(\rr)]^2 } & = & 4 f''(0) & +{\cal
  O}(N^{-1}); \nonumber \\
  N\cdot \ovl{ [\partial_{ii}V(\rr)]^2 } & = & 12 f''(0) & +{\cal
  O}(N^{-1}); \nonumber \\
  N\cdot \ovl{ \partial_{ii}V(\rr)\cdot\partial_{jj}V(\rr)  } & = &
  4 f''(0) & +{\cal  O}(N^{-1}); \nonumber \\
  \phantom{N\cdot} \ovl{ V(\rr)\cdot\partial_{ii}V(\rr) } & = & 
  2f'(0) & +{\cal  O}(N^{-1}); \nonumber \\
  N^{-1}\cdot \ovl{ [V(\rr)]^2 } & = & f(0) & +{\cal  O}(N^{-1}). 
\end{array}
\end{equation}
$\HH$ is split into a scalar part $\DD \delta_{ij}$ and a fluctuating
part $\HH'$. The elements of $\HH'$ are independent and gaussian, and
its eigenspectrum has, at the leading order, a semi-circular shape of
radius $4\sqrt{f''(0)}$ centred around 0.  If $N\to \infty$ and
$V(\rr)/N$ finite, then $\DD$ is constant, up to fluctuations of order
$N^{-1/2}$ (cf appendix~\ref{app:two}).
\begin{equation}
\label{eq:shift:H-diag}
  \DD = \frac{2f'(0)}{f(0)}\cdot \frac{V(\rr)}{N} + {\cal
  O}(N^{-1/2}).
\end{equation}
The resulting spectrum is the one announced in
equations~(\ref{eq:width:hessian}, \ref{eq:shift:hessian}).

In order to bridge the gap between (\ref{eq:slice-nm-definition}) and
(\ref{eq:1point:annealed}), we consider now the two-points annealed
average:
\begin{equation}
\label{eq:2points:annealed}
  \ovl{ \delta(V(\rr)-V_0)\ \delta(V(\rr_1)-V_1)\ \Sp(\HH)(\rr) }.
\end{equation}
The analysis involves now correlations between $V(\rr),
\partial_{ij}V(\rr), V(\rr_1), \partial_{ij}V(\rr_1)$. One finds that
for a generic correlator $f(y)$, $\Sp(\HH)(\rr)$ depends on both
$V(\rr)$ and $V(\rr_1)$. However, if $f(y)$ obeys $f\cdot f'' - (f')^2
= 0$, with $f(y)=\exp(-y)$ as a particular case, the dependence in
$V(\rr_1)$ disappears, and the result~(\ref{eq:shift:hessian}) holds.

Computing 
\begin{equation}
\label{eq:npoints:annealed}
\begin{array}{c}
  \ovl{ \delta(V(\rr)-V_0)\ \delta(V(\rr_1)-V_1) \ldots} \\
  \;\;\;\;\; \ovl{\times \delta(V(\rr_n)-V_n)\Sp(\HH)(\rr)} , 
\end{array}
\end{equation}
becomes very difficult as $n\geq 3$, and we were not able to find a
close expression for $\Sp(\HH)(\rr)\ (V_0, V_1,\ldots, V_n)$. However,
if $f\cdot f'' - (f')^2 = 0$, again $\Sp(\HH)(\rr)$ depends only on
$V_0$, and (\ref{eq:shift:hessian}) is valid.  This shows that the
spectrum of $\HH$ is a local quantity, independent of the environment
of the particle.

Because $\Sp(\HH(r))$ is a function of $V(\rr)$ only, we conclude that
the average~(\ref{eq:slice-nm-definition}) is described by
(\ref{eq:width:hessian}, \ref{eq:shift:hessian}) and that the
self-averaging property of $\Sp(\HH)(\rr)$ and its linear dependence
in $V(\rr)$, which was true for the $p$-spin model, is still true for
exponential correlators.  The appendix~\ref{app:two} gives further
details on the computation of (\ref{eq:1point:annealed}) and
(\ref{eq:2points:annealed}).

Now, we suppose that the trajectory ${\bf x}(t)$ explores
representative regions of the potential ({\it i.e.}  non-exceptional
points), for which the above mentioned results hold. The lowest
eigenvalue $-\SS$ of the hessian, defined by~(\ref{eq:shift:hessian})
becomes a time-dependent function:
\begin{equation}
  \SS(t) = \Lambda -\frac{2f'(0)}{f(0)} {\cal E}(t),
	\label{eq:time-dep:shift}
\end{equation}
leading to the energy dependent (through $\SS$) density of eigenvalues
of $\HH_{ij}$.  The number of eigenvalues between $\lambda-\SS(t)$ and
$\lambda-\SS(t)+\dd\lambda$ is $\rho(\lambda)\dd \lambda$ (time
independent).
\begin{equation}
  \rho(\lambda)  = 2(\pi\Lambda^2)^{-1}
  \sqrt{\lambda\ (2\Lambda-\lambda)}. 
        \label{eq:time-dep:spectrum}
\end{equation}
The marginality condition, by definition, is $\SS \equiv 0$.
Equation~(\ref{eq:time-dep:shift}) yields the ``geometrical energy'',
necessary for ${\cal H}$ to be marginal:
\begin{equation} 
\label{eq:energy-geom}
  {\cal E}_{geom} =  2 \frac{\sqrt{f''(0)} f(0)}{f'(0)},
\end{equation}
and the curvature ${\cal M}_{geom}$:
\begin{eqnarray}
\label{eq:curvature-geom}
  {\cal M}_{geom} & = & \int \!\dd \lambda \lambda \
  \rho(\lambda), \nonumber \\ 
  & = & 4 \sqrt{f''(0)}.
\end{eqnarray}
After a time $t$ long enough, the particle evolves in a marginal
region ($\SS(t)\simeq 0$) of the potential $V({\bf x})$, with a small
gradient $\|{\nab} V({\xx})\|$. At low temperature, the potential may
be developed up to the second order by means of local coordinates
$y_i$: $V(\yy) = V(0)+ \yy\cdot {\nab}V(0) + \sum_{i=1,N} \lambda_i
y_i^2/2$.  The plateau value ``$q$'' of the correlation function
$b(t,t')$ is thus given by assuming that each direction of curvature
$\lambda_i$ is thermalized with $\mean{ y_i^2} \simeq T/\lambda_i$,
and $q=2N^{-1}\sum_{i=1,N} \mean{y_i^2}$:
\begin{eqnarray}
 \qb_{geom}  & = & \int_0^{8\sqrt{f''(0)}} \!\dd \lambda \frac{2T}{\lambda} 
 \ \rho(\lambda)|_{S=0}, \nonumber \\
 & = & \frac{T}{\sqrt{f''(0)}}.
\end{eqnarray}
Let us compare now with the results from the dynamical mean-field
theory, in the zero temperature limit~\cite{CugLeD}.
\begin{eqnarray}
\label{eq:exactvalues}
  \lim_{t \to \infty} {\cal E}(t) & = & \frac{f'(0)}{\sqrt{f''(0)}} +
  \frac{f(0) \sqrt{f''(0)}}{f'(0)},  \\
  \lim_{t \to \infty} {\cal M}(t) & = & 4 \sqrt{f''(0)}, \\
  \qb  & = & \frac{T}{\sqrt{f''(0)}}.
\end{eqnarray}
Agreement holds for the curvature and $\qb$, whereas the geometrical
and dynamical energy differ, unless $f(0)f''(0)=f'(0)^2$.  We cannot
conclude about the relevance of the geometrical approach for a generic
correlator, \textit{e.g.} power law, as (\ref{eq:shift:hessian})
probably does not hold.  However, the exponentially correlated
toy-model turns out to be a very favourable model, for which the
geometrical approach gives reasonable results. The following of this
paper aims at demonstrating that many features of the zero temperature
dynamics of this model (exponents, aging, driving with a force) can be
explained with the help of geometrical arguments.


\section{The distribution of the gradient's coordinates}
\label{sec:five}

In this  section, we define  an orthonormal frame ``attached''  to the
particle. The procedure used is reminiscent from the definition of the
instantaneous  normal  modes  in  the  study  of  supercooled  liquids
dynamics~\cite{Keyes}. Then, we investigate the statistical properties
of the components  of $\nab V(\xx(t))$ in this  special frame. We find
that  these components  are  distributed according  to a  self-similar
form, determined by the value of the exponent~$\kappa$ of the energy
decay. 

We develop up to the second order the potential around the actual
position of the particle $\xx(t)$.
\begin{eqnarray}
  {\bf Q}(\xx)  & = &  V(\xx(t)) + \sum_i (\xx_i-\xx_i(t))
  \cdot\partial_i V(\xx(t)) \nonumber\\
  & & \hspace{-1.5cm} + 1/2\sum_{ij} \partial_{ij} V(\xx(t))\cdot
  (\xx_i-\xx_i(t))\cdot(\xx_j-\xx_j(t)), 
\label{eq:local-environment}
\end{eqnarray}
We define an orthonormal frame of eigendirections $\{ {\bf
e}_{\lambda_i}(t) \}$ in which the hessian ${\cal H}_{ij}(t) =
\partial_i \partial_j V(\xx(t))$ is diagonal.  $\lambda_i$ belongs to
the --time independent-- interval $[0,2\Lambda]$, so that the
corresponding eigenvalue of $\mathbf{Q}$ is just $\lambda_i-\SS(t)$.

We follow ``adiabatically'' the eigenvectors $\{ {\bf
e}_{\lambda_i}(t) \}$ as the particle moves. A mild assumption is that
the $\{ {\bf e}_{\lambda_i}\}$ evolve smoothly, provided the levels
$\lambda_i$ are allowed to freely cross each other. This choice
implies that any ordering of the $\lambda_i$ lasts only for a short
period of time. The $\{ {\bf e}_{\lambda_i}\}$ define a comoving
frame, in which the gradient $\nab V$, or equivalently the velocity,
can be projected.
\begin{eqnarray}
  -{ \nab} V(\xx(t)) & = & \sum_{i} {\gam}_i(t) \cdot
  \ee_{\lambda_i}(t), \\
  & = & \dot{\xx}(t). \nonumber
\end{eqnarray}
There are reasons to consider that the components $\gam_i(t)$ are
randomly and evenly distributed, even in the deterministic zero
temperature limit. First, this randomness reflects the average over
the ``white'' initial conditions. Then, as the correlator $\overline{
\partial^2_{ij} V(\xx) \partial^2_{ij} V(\xx')}$ is exponentially
short range correlated, one can suppose that the comoving frame is
rotating on itself in a chaotic manner, as it does in the spherical
($p \geq 3$)-spin model~\cite{KurLal}. So, during the particle's
motion, each component $\gam_i$ spreads continuously over the $N-1$
others directions.

The sign of $\gam_i(t)$ itself is irrelevant, because of the arbitrary
definition of the frame, invariant under the reflections
$\ee_{\lambda_i} \leftrightarrow -\ee_{\lambda_i}$.  We claim that
$\gam_i^2(t)$ has to be preferred to $\gam_i(t)$.  On physical
grounds, we propose to consider only the smoothed quantity
$\locav{\gam_i^2(t)}$, obtained by averaging locally over the few
$\sqrt{N}$ indices $j$ such that $\lambda_i -N^{-1/2} < \lambda_j <
\lambda_i + N^{1/2}$. This is possible because the mean interspacing
between the $\lambda_i$ is ${\cal O}(N^{-1})$. As $N$ goes to
$\infty$, one expects $\locav{\gam_i^2(t)}$ to become a smooth
function of $\lambda_i$, varying only on the scale $\delta \lambda
\sim 1$ (although, rigorously, the scale of variation is $\delta
\lambda \sim N^{-1/2}$), making the dependence in the index~$i$
irrelevant.

The function:
\begin{equation}
  g(\lambda_i,t) = \locav{\gam_i^2(t)},
  \label{eq:local-av}
\end{equation}
is the distribution of the gradient's coordinates (or equivalently of
the instantaneous velocity coordinates) and is a central object in the
present study.
 
In this continuous limit, the two first time derivatives of $\EE$ can
be expressed with the help of the density $\rho(\lambda)$ and the
distribution $g(\lambda,t)$ as:
\begin{eqnarray}
  \dot{{\cal E}}(t) & = & - \sum_i \partial_i V(\xx(t)) \cdot
  \partial_i V(\xx(t)), \nonumber \\ 
  & = & -\int \!\dd \lambda \ \rho(\lambda)\ g(\lambda,t).  
  \label{eq:E-dot} \\
  \ddot{{\cal E}}(t) & = & \sum_{ij} \partial_j V(\xx(t)) \cdot
  \partial_{ij} V(\xx(t)) \cdot  \partial_i V(\xx(t)),
  \nonumber\\  
  & = & \int \! \dd \lambda\ \rho( \lambda)\ (\lambda - \SS(t))\
  g(\lambda,t). \label{eq:E-ddot} 
\end{eqnarray}
In~\cite{KurLal} was already noticed that, due to the algebraic
decay of the energy ${\cal E}(t)=-2+1.08\cdot t^{-\kappa}$, the ratio
$\ddot {\cal E}(t)/\dot {\cal E}(t)$ was $\sim 1/t$.  From
section~\ref{sec:three}, we know that $\SS(t)\sim t^{-0.67}$, which
implies $\ddot{\cal E}(t) \ll \SS(t)\cdot \dot{\cal E}(t)$, and thus:
\begin{eqnarray}
  \int\! \dd \lambda\  \rho(\lambda)\; \lambda \cdot g(\lambda,t)  
  & = & \ddot {{\cal E}}(t) + \SS(t) \int \!\dd \lambda\ \rho( \lambda)
  g(\lambda,t), \\
  & \simeq &   \SS(t) \int \!\dd \lambda\ \rho( \lambda) g(\lambda,t).
  \label{eq:momentofg}
\end{eqnarray}
The first moment of $g(\lambda,t)\ \rho(\lambda)$, is proportional to
$\SS(t)$, suggesting a self-similar scaling form for $g(\lambda,t)$,
valid for $t\to \infty$ and $T=0$ (Figure~\ref{fig:gradient}):
\begin{equation}
  g(\lambda,t) = \Gamma(t)\  \hat{G} \left( \frac{\lambda}{\SS(t)}
  \right),  \label{eq:scaling-form}
\end{equation}
The knowledge of the other moments of $g$ would be useful to confirm
equation~(\ref{eq:scaling-form}), but unfortunately, they are very
difficult to compute, and are no more given by the next derivatives of
${\cal E}$. 

As $t$ increases, only the smaller $\lambda_i$ keep any relevance, and
the density $\rho$ is well approximated by its $\lambda\sim 0$
equivalent $\pi^{-1}(2/\Lambda)^{3/2}\sqrt{\lambda}$.  In this limit,
the loss of energy rate becomes, from (\ref{eq:E-dot}) and
(\ref{eq:scaling-form}):
\begin{equation}
  \dot{\SS}(t) = -\frac{2f'(0)}{f(0)} \dot{\EE}(t) \propto 
  -\Gamma(t)\cdot \SS(t)^{3/2}. \label{eq:loss-energy}
\end{equation}
The knowledge of the exponent $\kappa$ of $\SS(t) \sim t^{-\kappa}$
(section~\ref{sec:three}) fixes the prefactor $\Gamma$ up to a
constant, to:
\begin{equation}
  \label{eq:link-Gamma-S}
  \Gamma = \SS^{(2-\kappa)/2\kappa}.
\end{equation}
Our section~\ref{sec:two} suggests $\kappa$ is very close to $2/3$,
which would imply $\Gamma\propto \SS$.

The next momentum of $g(\lambda,t)\rho(\lambda)$ provides information
on the time correlations of the unit vector ${\bf w}(t)$ of the
particle's trajectory. One the one hand,
\begin{eqnarray}
\Big( \partial_t {\nab} V(\xx(t)) \Big)^2 
  & = & 
  \sum_i \left( \sum_j -\partial_j V(x(t)) \partial_{ij} V(x(t)) \right),
  \nonumber\\ 
  & & \times \left( \vphantom{\sum_j} \sum_k -\partial_k V(x(t))
  \partial_{ik}  V(x(t))\right), \nonumber \\
  & = & \sum_{i,j,k} \partial_j V \cdot \partial_{ji} V \cdot
  \partial_{ik} V  \cdot \partial_k V, \nonumber \\
  & = & \int \! \dd \lambda\  \rho(\lambda) \; (\lambda -\SS(t))^2 \;
  g(\lambda,t). 
\end{eqnarray}
With the scaling form for $g(\lambda,t)$, the right hand side is of
order $\Gamma(t) \SS(t)^{7/2}$. On the other hand, we perform a
decomposition $-\nab V(\xx(t)) = M(t)\ {\bf w}(t)$. The norm $M(t)$
equals $(- \dot{{\cal E}}(t))^{1/2}$, and ${\bf w}(t)$ is the unit
vector, tangent to the trajectory.  The following equality holds:
\begin{equation}
  \left( \partial_t {\nab} V \right)^2  = \left( \partial_t M \right)^2
  + M^2 \cdot \| \partial_t{\bf w}\|^2.  
\end{equation}
This sum is clearly dominated by $M^2\cdot\|\partial_t{\bf w}\|^2$,
with $M^{2} \simeq \Gamma\cdot \SS^{3/2}$. The unitary vector rotates,
regardless to the actual value of $\Gamma(t)$, at a rate $ \|
\partial_t{\bf w} \|\sim \SS(t)$.  One expects the ``director'' ${\bf
w}(t)$ to have changed its orientation after a typical time
$\SS(t)^{-1}$, which looks like a ``persistence time'' for the
trajectory of the particle.  Consequently, the motion of $\xx(t)$
crosses over from a ``ballistic'' regime $\| \xx(t+\delta t)-\xx(t) \|
\sim M^2\cdot (\delta t)^2;\ \delta t \ll \SS^{-1}$ to a diffusive
regime $\| \xx(t+\delta t)-\xx(t) \| \sim D\cdot \delta t;\ \delta t
\gg \SS^{-1}$. The existence of a diffusive regime is here inferred by
the expansion~(\ref{eq:scal-form-ageing}, \ref{eq:definition-tb}),
valid only for exponentially correlated disorder, and by no means
generic.


\section{A short time, quasi-static approximation}
\label{sec:six}

We investigate here the breakdown of the fluctuation dissipation
theorem, in the zero temperature limit. The fluctuation-dissipation
violation is measured by the function $\Tb(t,t')$ (cf
\ref{eq:qfdt:2}). We propose here a model for the short time
evolution of $\Tb(t,t')$, and show that its predictions are in good
agreement with the findings of section~\ref{sec:two}.

We approximate locally the potential $V(\xx)$ around $\xx(t)$ by a
quadratic function (\ref{eq:local-environment}), which may can be
considered as constant provided we restrict ourselves to a time
separation $t-t'$ small enough. One can always find a coordinate
system $\{ {\bf y}_i \}$ such that this quadratic potential reads:
\begin{equation}
  {\bf Q}(\yy) = {\bf Q}(0)+ 1/2 \sum_i
  (\lambda_i-\SS)\cdot y_i^2, 
  \label{eq:quad-pot}
\end{equation}
where the coordinates ${\bf y}$ must not be confused with the original
coordinates ${\bf x}$ of the relaxational motion.

This section aims at demonstrating that when a particle diffuses, or
relaxes in such a parabolic potential, then a characteristic time
$t_f$ scaling like $\SS^{-1}$ arises, which turns out to be the time
scale along which the function $\Tb(t,t')$ departs from its
equilibrium value~0, \textit{i.e} the fluctuation-dissipation
violation characteristic time.

We consider a particle moving on the potential (\ref{eq:quad-pot}),
starting at $t_0$, and define the time difference $\tau=t-t_0$.  The
intermediate steps of the calculation make use of $\tau,t_0$, while
the final results are expressed in term of $t,t'$ in relation with the
original out-of-equilibrium relaxation.

Let us consider the same local average $\locav{{\bf y}_i^2(\tau)}$ as
in equation~(\ref{eq:local-av}). The ${\bf y}_i$ are related to the
gradient's coordinates by $\gam_i(\tau)=-(\lambda_i-S)\cdot
y_i(\tau)$:
\begin{equation} 
  \locav{y_i^2(\tau)}= \frac{g(\lambda,t_0+\tau)}{(\lambda-\SS)^2}.
\end{equation}
The initial conditions $\locav{y_i^2(\tau=0)}$ are given by
$g(\lambda,t_0)$.  One computes the fluctuation dissipation
violation $\Tb(t,t')$, when the quadratic potential
(\ref{eq:quad-pot}) does not evolve with time ($\SS$ fixed once for
all), and with initial conditions  arising from a
realistic distribution $g(\lambda,t_0) = \Gamma\cdot
\hat{G}(\lambda/\SS)$.
\begin{equation}
  y_i(\tau)= y_i(0)\ e^{-(\lambda_i-\SS)\cdot(\tau)}. \label{eq:relax:saddle}
\end{equation}
The distribution $g(\lambda,t_0+\tau)$ evolves like
$\locav{{\bf y}_i^2(\tau)}\cdot (\lambda-\SS)^2$. One has, far all
$t>t_0$: 
\begin{eqnarray}
  g(\lambda,t_0+\tau) & = & g(\lambda,t_0)\ e^{-2(\lambda-\SS)\cdot(\tau)};
  \nonumber\\
  \partial_{t}\ g(\lambda,t) & = &  -2g(\lambda,t)\cdot(\lambda-\SS).
  \label{eq:partial-g}
\end{eqnarray}
The usual response $r(t,t') = N^{-1} \sum_i \delta y_i(t)/\delta
\zeta_i(t')$, and correlation $b(t,t') = N^{-1} \sum_i
(y_i(t)-y_i(t'))^2$ functions reexpress in terms of $g(\lambda,t')$:

\begin{eqnarray}
  r(t,t') & = & \int \dd \lambda\ \rho(\lambda)\
  e^{-(\lambda-\SS)\cdot(t-t')}; \label{eq:r-g}\\
  b(t,t') & = &  \int \dd \lambda\ \rho(\lambda)  
  g(\lambda,t')\cdot \left( \frac{1-e^{-(\lambda-\SS)\cdot
  (t-t')}}{\lambda-\SS} \right)^2; \label{eq:b-g}\\
  \partial_{t'} b(t,t') & = & -2 \int \dd
  \lambda\ \rho(\lambda)\ g(\lambda,t')
  \left(\frac{1-e^{-(\lambda-\SS)\cdot(t-t')}}{\lambda-\SS}\right); 
                     \nonumber\\
	& = & -2 \Tb(t,t'). \label{eq:partial-b-g} 
\end{eqnarray}

By inserting $g(\lambda,t'=t_0)=\Gamma \hat{G}(\lambda/\SS)$ in
(\ref{eq:partial-b-g}), one deduces the short time, $\tau \ll \SS^{-1}$,
value of $\Tb(t,t')$,
\begin{equation}
  \Tb(t,t') = (t-t')\cdot \Gamma \cdot \SS^{3/2},
\end{equation}
and the intermediate time $\tau \sim \SS^{-1}$ one,
\begin{eqnarray}
  r(t,t') & = & \SS^{3/2} \Phi_0( \SS\cdot(t-t')), \nonumber\\
  b(t,t') & = & -2 \Gamma\cdot \SS^{1/2}\cdot \Phi_1(\SS\cdot(t-t')),
  \nonumber\\ 
  \Tb(t,t') & = & \frac{\Gamma}{\SS} \frac{\Phi_0}{\Phi_1}(\SS\cdot
  (t-t')). \label{eq:short-time-fdt-violation}
\end{eqnarray}
$\Phi_0$,$\Phi_1$ are scaling function presented in
appendix~\ref{app:two}. Equation~(\ref{eq:short-time-fdt-violation})
shows that $\SS^{-1}$ plays the role of a characteristic time for the
onset of the effective temperature $\Tb$.

Assuming now the very likely value $\kappa=2/3$ and $\Gamma=\SS$, one
finds that $\Tb$ becomes an order one quantity after a time $t_f \sim
\SS^{-1}$. One concludes that the characteristic time scale $t_f$ should
scale like $t_f = t^{\alpha} \propto t^{\kappa} $, and
$\alpha=\kappa=2/3$.

Our numerics (Figure~\ref{fig:tf}) lead to an estimated value $\alpha
\simeq 0.64$, while $\kappa \simeq 0.67$. While we haven't proved that
$\kappa$ is actually $2/3$, we find the agreement satisfactory, and
believe that the above picture describes correctly the first stage of
the breaking of the ``fluctuation-dissipation relation at zero
temperature''.

How long can the quadratic approximation~(\ref{eq:local-environment})
accurately describe the original relaxational process ?  As $\SS(t)$
decreases algebraically, the necessary time $\delta t$ to have
$|\SS(t+\delta t)-\SS(t)| \sim \SS(t)$ is $t$ itself.  More seriously,
we have seen in the previous section, that the unit vector of the
trajectory $\dot{{\bf x}}(t)$ changes with the time scale
$\SS^{-1}\sim t^{2/3}$. As this change is somewhat related to the
frame's chaotic motion, we deduce than $\SS^{-1}$ must be an upper
limit of validity of the quasi-static approximation.  Finally, the
relaxation on the saddle becomes ill-defined when $\SS\cdot (t-t') \gg
1$, due to the exponential divergence of the functions $r(t,t')$ and
$b(t,t')$, given by the equations~(\ref{eq:r-g}, \ref{eq:b-g}). We
arrive to the conclusion that this quasi-static picture is not valid
beyond times much greater than $\SS^{-1}$, but provides a strong
presumption in favour of $t_f(t) \sim \SS^{-1}(t)\sim t^{2/3}$, in
good agreement with our numerical findings (section~\ref{sec:three}
and Figure~\ref{fig:tf}).

Let us close this section by computing the typical distance covered
during a time interval $t-t' < \SS^{-1}$, with a gradient coordinates
distribution $g(\lambda,t') = \Gamma \cdot \hat{G}(\lambda/\SS)$.
\begin{eqnarray}
  b(t,t') & = &  \Gamma \SS^{3/2} \cdot (t-t')^2 \nonumber\\
   & = & \SS^{5/2} \cdot (t-t')^2\ \mbox{if}\ \kappa=2/3.
  \label{eq:balistic-step}
\end{eqnarray}
For a time interval $(t-t')\sim \SS^{-1}$, $ b(t,t') \sim
\Gamma\cdot\SS^{-1/2} \ll 1$, becoming $b(t,t') \sim \SS^{1/2}$ for
$\kappa=2/3$. As $\SS^{1/2}$ tends to zero, the characteristic time
$t_b$ of the evolution of $b(t,t')$ in the aging regime, is
necessarily much greater than $t_f \sim \SS^{-1}$.


\section{A dynamics restricted to the downhill directions}
\label{sec:seven}

This section shows how the above approach describes the long time
aging regime.

The equation~(\ref{eq:loss-energy}) has a simple physical
interpretation. The only non-vanishingly small components $\gamma_i$ of $-\nab
V$ are those corresponding to $\lambda_i \leq \SS$. Only a number
$N\int_0^{\SS} \dd\lambda \rho(\lambda) \sim N\cdot \SS^{3/2}$
directions $i$ are contributing to $-(\nab V)^2 = -\sum_i
\gam_i^2$. Each one of these $\gam_i$ has a magnitude of order
$\gam_i^2 \sim \Gamma$. As a result, $N\cdot \dot{\EE}$ scales like
$-\sum_i \gam_i^2 = N\cdot \SS^{3/2}\cdot \Gamma$.
\begin{equation}
  \dot{\EE} \propto -(\EE(t)-\EE_d)^{1+1/\kappa}.
\end{equation}
The relaxation dynamics looks like if it was controlled by the difference
$\EE(t)-\EE_d$.

\textbf{The linear response regime.} A constant force $\FF$ is now
applied, uncorrelated to the potential $V$. Each one of its (comoving)
coordinate $f_i$ is random, time-dependent as the frame rotates during
the particle's motion, and has a magnitude $f_i\sim F$.  We suppose
$F$ weak enough to be considered as a perturbation around the
relaxational dynamics described in section~\ref{sec:six}.

At any time, there are ``open'', or downhill directions, with
$\lambda_i\leq \SS$ and ``close'', or uphill directions with $\lambda_i
\geq \SS$. The close directions behave as confining harmonic potentials
which prevent the (weak) force $f_i$ to drive the particle along this
direction. The open directions are the one along with the external
force drives efficiently the particle away. As the particle moves,
``open'' and ``close'' directions exchange their role, but the
proportion of open directions remains proportional to $\SS^{3/2}$.

The force $\FF$ induces a displacement $\dot{\xx}$ whose components
are $\dot{\xx}_i \simeq f_i$ along an open direction and
$\dot{\xx}_i\simeq 0$ along a close direction . The average velocity
$\dot{\xx}\cdot\FF/ \| \FF \| $ is given by ($\theta$ Heaviside
function):
\begin{eqnarray}
 \frac{\dot{\xx}(t)\cdot \FF}{ \| \FF \|} & \simeq &
 \sum_i f_i^2\ \theta(\SS-\lambda_i)/ \| \FF \| ; \nonumber\\
  & = &  N\cdot \SS^{3/2}\cdot F^2/ (\sqrt{N} F); \nonumber\\
  & = & \sqrt{N}\ F\ \SS^{3/2}; 
  \label{eq:linear-response:F}
\end{eqnarray}
that we identify to $\sqrt{N}\ \dot{u}(t)$. As a result, one finds a
velocity proportional to the number of downhill directions:
\begin{equation}
  \dot{u} \propto F\cdot \SS^{3/2}.
\end{equation}
Inserting the likely value $\kappa=2/3$, one finally gets a
displacement $u(t)-u(t') \propto F (\ln(t)-\ln(t'))$, well confirmed
by the numerics (section~\ref{sec:three} and
Figure~\ref{fig:log-disp}).  This pure relaxational motion is driven by
the components $\gam_i$ of $-\nab V$ along the open directions, while
the external force acts with $f_i$ along the same open directions. One
expects the linear response to hold if $f_i^2 \ll \gam_i^2$ but to
break down when $f_i^2 \simeq \gam_i^2$. This leads to a predicted
cross-over time scaling like $\Gamma(t_F) = F^2$, or $t_F\sim
F^{4/(\kappa-1)}$, to be investigated in a forthcoming
publication~\cite{ComingSoon}.

\textbf{The diffusive regime.}
The asymptotic behaviour predicted for $b(t,t')$ is, from equations
(\ref{eq:scal-form-ageing}-\ref{eq:definition-tb}):
\begin{equation} 
  \label{eq:diffusive-behavior}
    b(t,t') = \frac{t-t'}{t_b} + {\cal O} \left( \frac{t-t'}{t_b}
  \right)^2.
\end{equation}
One recognises a simple diffusive behaviour, with effective
diffusivity $t_b^{-1}$. From section~\ref{sec:four}, we know that the
short-time motion $(t-t') \leq \SS^{-1} \sim t_f$ is ballistic, and
that the particles covers a distance of $\Gamma\cdot \SS^{-1/2}$ On
the other hand, our results from section~\ref{sec:three} show that the
direction ${\bf w}$ of the trajectory $\xx(t)$ uncorrelates itself
after this same time $\SS^{-1}$. Using a well-known result on
correlated random walks, and assuming a free diffusive behaviour at
intermediate times $t-t'\sim t_b$, as inferred by
equation~(\ref{eq:diffusive-behavior}), one finds:
\begin{equation}
  \label{eq:correlated-ran-walk}
  b(t,t') \simeq \left( \frac{t-t'}{\SS^{-1}} \right)\times
  \left(\frac{\Gamma}{\sqrt{\SS}}\right). 
\end{equation}
This corresponds to ballistic steps of length $(\Gamma/\sqrt{\SS})$
(\ref{eq:balistic-step}), and a cross-over time from ballistic to
diffusive regime equal to $\SS^{-1}$.  As $t_f\simeq\SS^{-1}$,
\begin{equation}
  \label{eq:begining:aging-regime}
  b(t,t') \simeq (t-t')\cdot \Gamma\cdot\SS^{1/2},
\end{equation}
leading to the identification:
\begin{eqnarray}
  t_b & \propto & \Gamma^{-1}\SS^{-1/2}; \nonumber\\
      & \propto & \SS^{1/\kappa} \label{eq:characteristic-time:aging}
\end{eqnarray}
If $\kappa$ is taken to be $2/3$, one gets $t_b \sim t' \sim t$. 

Let us discuss different reasons to be confident in the scaling $t_f
\sim \SS^{-1}$, $t_b \sim \SS^{-3/2}$ and $\SS \sim t^{-2/3}$.  First,
a matching argument similar to~\cite{KinHor:1,Horner} predicts $t_b
\sim t_f^{3/2}$. Then, the result $t_b \sim t'$ is in agreement with
the conjecture $h(t) \sim t^{\delta}$. This entails a logarithmic
growth of $b(t,t')$, $t'$ fixed, and we have asymptotically ({\it
i.e.} $t,t' \gg 1$, and $t/t'\sim 1$) a free brownian motion in
logarithmic time.
\begin{equation}
  \label{eq:brownian:log-time}
   b(t,t') = \delta\cdot (\ln t - \ln t').
\end{equation}
This makes $\exp(-b(t,t'))$ as well as $r(t,t')$ decaying as a power
law. While we have no demonstration of that, we think that a power-law
decay of the memory function $f(b(t,t'))r(t,t')$ is necessary for the
``fine tuned'' aging solution of the system~(\ref{eq:motion:equations}).
Asking for a power-law decay $f(b(t,t'))$ in turn fixes $\kappa$ to
$2/3$.

Finally, if $\kappa=2/3$, $t_b=\SS^{-3/2}$, and the characteristic times
for the linear response regime $\dot{u}(t) \simeq F/t_b$, and for the
diffusion regime $b(t,t') \simeq (t-t')/t_b$ are the same, which is
consistent with the persistence of an ``Einstein relation'' at the
beginning of the aging regime.


\section{Conclusion}
\label{sec:nine}

We have proposed a geometrical description of the mean-field
relaxational dynamics of a particle, for a subclass of short-range
correlated disorders. We have restricted ourselves to the isolated
case, and to the driven case in the linear response regime.

A numerical integration of the mean-field equations gives evidence of
a power-law decay of the dynamical energy with an exponent $\kappa$
numerically close to $2/3$. We also found evidence of a logarithmic
growth $b(t,t') \sim \ln t$ consistent with the conjecture $h(t) \sim
t^{\delta}$ for the reparametrization function $h$.

The exponential correlator makes it possible to compute the density of
eigenvalues of the hessian $\HH$ associated to the random potential,
and we were able to predict the correct value ({\it i.e.} $-2$) of the
dynamical energy $\EE_d$.
 Introducing a comoving frame, reminiscent from the
INM frame of a supercooled liquid, we derive an expression for the
distribution $g(\lambda,t)$ of the components of $\nab
V\big(\xx(t)\big)$. This expression is $g\ = \Gamma\
\hat{G}(\lambda/\SS)$, where $-\SS(t)$ is the (time dependent) lowest
eigenvalue of $\Sp(\HH)$.

For reasons exposed in section~\ref{sec:seven}, namely the consistence
with $h(t) \sim t^{\delta}$, the requirement that $f(b)$ is likely to
decrease as a power law, and acknowledging the numerical estimate of
$\kappa$, we believe that $\kappa$ is indeed equal to $2/3$. This
leads to the following predictions:\\
\indent {\it (1)} $\Gamma \propto S$, and a typical gradient
coordinate is, along a downhill direction, $|\gam_i|\sim \sqrt{S} \sim
t^{-1/3}$.\\
\indent {\it (2)} From a short time, harmonic expansion of the
particle's motion (section~\ref{sec:six}) the characteristic time
$t_f$ leading to the appearance of an effective temperature goes like
$t_f \sim S^{-1} \sim t^{2/3}$.\\ 
\indent {\it (3)} The characteristic time at the beginning of the
aging regime is $t_b \sim S^{-3/2}\sim t$. Both linear response
$\dot{u} \sim F/t_b$ and diffusion $b(t,t') \sim (t-t')/t_b$ are
controlled by it.\\

We conclude that the aging mechanism of this model comes from a
simultaneous decrease of the number of downhill directions (going like
$N\cdot S^{3/2} \sim Nt^{-1}$) and of the typical gradient component
$|\gam_i|\sim t^{-1/3}$.

We predict also that the effect of a constant force brings about a
dramatic change in the dynamics after a time $t_F \sim F^{-3}$,
reaching a out-of-equilibrium but stationary regime~\cite{ComingSoon}.

\paragraph*{Acknowledgements} I especially thank L.Cugliandolo and
J.Kurchan for having lent me their numerical code, and S.Scheidl, J.P
Bouchaud, J.Kurchan, M.M\'ezard and A.Cavagna for discussions on this
field.  I thank D.Feinberg for suggestions and criticisms about the
manuscript.  I warmly thank the hospitality of the Department of
Physics, IISc, Bangalore, where a part of the writing has been done.

\appendix
\section{The MSR action}
\label{app:one}
The action leading to equations~(\ref{eq:motion:equations}) is:
\begin{eqnarray}
   S[x,i\tx] & = & \int_0^{\infty}\dd t\ \left\lbrace 
     -T\sum_{j=1}^{N} (i\tx_j)^2(t) + i\tx_1(t) \cdot
      (\dot{x}_1(t)-N^{1/2}\cdot F) + \sum_{j=2}^{N} i\tx_j(t) \cdot
      \dot{x}_j(t) \right\rbrace \nonumber\\
            &   & + \int_{0}^{\infty} \dd t\ \dd s\ 
     \left\lbrace f'(d(t,s))\sum_{j=1}^{N} i\tx_j(t)\cdot i\tx_j(s) 
     + 4f''(d(t,s))\ r(t,s)\ \sum_{j=1}^{N}i\tx_j(t)\cdot (x_j(t)-x_j(s))
     \right\rbrace,  \label{eq:gaussian-action}
\end{eqnarray}
and the expectation value of an observable ${\cal
O}[x(t),i\tilde{x}(t)]$, averaged over the disorder, is given by:
\begin{equation}
  \ovl{\mean{{\cal O}}} = \int {\cal D}x[t]\ {\cal D}\tilde{x}[t]\ {\cal O}\
  \exp(-S).
\end{equation}
%


\section{The spectrum of the Hessian ${\cal H}$}
\label{app:two}

We consider $\ovl{ \delta(V(\rr)-V_0)\ \Sp(\HH)(\rr) }$ for any
arbitrary potential. $V_0 = N\EE$ is fixed, and the $\HH_{ij} =
\partial^2_{ij} V(\rr)$ are $N(N+1)/2$ gaussian random variables. The
correlations among the $\HH_{ij}$ are listed in
equation~(\ref{eq:correlations:V}). We define the self-averaging
quantity $\DD = N^{-1}\sum_i \partial_{ii}V(\rr)$ so that:
\begin{eqnarray}
N \ovl{ (\partial_{ii} V(\rr) - \DD)^2} & = & \; 8f''(0)-8f''(0)/N,
\nonumber\\ 
N \ovl{ (\partial_{ii} V(\rr) - \DD) V(\rr) } & = & \; 0,  \nonumber\\
N \ovl{ (\partial_{ii} V(\rr) - \DD)(\partial_{jj} V(\rr) - \DD)}
  & = &  -8f''(0)/N, \nonumber\\
  & = & \; 0 + {\cal O}(N^{-1}). 
\end{eqnarray}
The hessian is now $\HH_{ij}= \DD \delta_{ij} + \HH'_{ij}$.  $\HH'$ is
a matrix of independent gaussian centred random numbers. The diagonal
elements are slightly correlated (of order $1/N^2$) and have a
different variance than the off-diagonal elements, but this does not
prevent the Wigner result to apply and the spectrum of $\HH'$ is a
centred semi-circle of width $\Lambda=4\sqrt{f''(0)}$.

The determination of $\DD$ follows from the fact that $\EE$ and $\DD$
are gaussian distributed, with correlations:
\begin{eqnarray}
N\ovl{\EE^2} & = & f(0); \nonumber\\
N\ovl{\DD^2} & = & 4\left( \frac{N+2}{N} \right) f''(0);\nonumber\\
N\ovl{\DD\cdot\EE} & = & 2f'(0). 
\end{eqnarray}
The joint probability distribution of $\EE$ and $\DD$ is:
\begin{eqnarray}
{\cal P}(\DD,\EE) & = & \frac{N}{2\pi\sqrt{c f(0)}}\exp\left(
-\frac{N}{2}\left[ \frac{\EE^2}{f(0)} + \frac{(\DD-a\EE)^2}{c} \right]
\right); \nonumber\\ 
c & = & \frac{4}{f(0)}\cdot (f(0)f''(0) -f'(0)^2) + \frac{8}{N} f''(0);
\nonumber\\
a & = & \frac{2f'(0)}{f(0)}.
\end{eqnarray}
Fluctuations of $\DD$ are of order $N^{1/2}$ around its saddle point
value $2f'(0)/f(0)\times\EE$. It follows that $\Sp(\HH)$ is a semi-circle
of radius $\Lambda$ shifted by an amount $\DD= 2f'(0)/f(0)\times\EE$.

Let us consider now\\ $\ovl{ \delta(V(\rr)-V_0)\ \delta(V(\rr_1)-V_1)\
\Sp(\HH)(\rr) }$. This simple average measures the non-locality of
$\Sp(\HH)$, {\it i.e.} its dependence on the values taken by the
random potential $V(\rr')$ around $\rr$. 

The rotational invariance of the above average is broken, and $\rr -
\rr'$ plays a special role. We relabel hereafter the direction~1 to
coincide with $\rr-\rr'$, and define
$b=\|\rr-\rr'\|^2/N$. Correlations are now, in addition
to~(\ref{eq:correlations:V}),
\begin{eqnarray}
\ovl{ \partial_{ii} V(\rr) V(\rr')} & = & 2 f'(b)\ \mbox{if}\ i\geq 2;
\nonumber\\
\ovl{ \partial_{11} V(\rr) V(\rr')} & = &  2 f'(b)+4b
f''(b);\nonumber\\ 
\ovl{ \partial_{ij} V(\rr) V(\rr')} & = & 0. \label{eq:more-correlations}
\end{eqnarray}

$\DD$ is again defined as $N^{-1}\sum_i \partial_{ii} V(\rr)$ and
$\HH'_{(i,j)\geq 2}$ is equivalent to the above situation:
independent, centred, gaussian random components, and the spectrum is
a centred semi-circle. Adding one row and one column of random
independents elements to $\HH'_{(i,j) \geq 2}$ must not change the
density profile of eigenvalues. This is because this eigenvalue
distribution is a fixed point under the change $N \to N\!+\!1$, as
argued in the cavity approach of the problem. A possible trouble come
from the single component $\partial_{11}V-\DD$ which does not average
to 0, but this does not alter the final result more than by a single
isolated eigenvalue. 

It is possible to show, with the help of a formal field theoretical
approach, that the
correlations~(\ref{eq:correlations:V}),(\ref{eq:more-correlations})
indeed lead to the ordinary $N\to \infty$ saddle point for
$\Sp(\HH')$, {\it i.e.} a semi-circle law of radius~$\Lambda$.

The computation of $\DD$ follows closely the lines of the previous
paragraph. We found that if $\EE= V(\rr)/N$, $\EE'= V(\rr')/N$ and 
$b=\|\rr-\rr'\|^2/N$, then:
\begin{eqnarray}
\DD & = & \left( \frac{ f'(0)+f'(b)}{f(0)+f(b)} \right)\cdot
(\EE+\EE') \nonumber\\
  &  & +  \left( \frac{ f'(0)-f'(b)}{f(0)-f(b)} \right)\cdot
(\EE-\EE').
\end{eqnarray}
For a generic correlator, there is an explicit dependence on $\EE'$
(``non locality'') while for an exponential correlator $f =\exp (-y)$,
the above formula reduces to $\DD= 2f'(0)/f(0)\times\EE$.

This suggests that the determination of $\Sp(\HH)$ from
(\ref{eq:slice-nm-definition}) is a complex problem and the simple
behaviour~(\ref{eq:shift:hessian}) fails for a generic~$f$.

The exponential correlator, however, has a strong property. The
average:
\begin{eqnarray}
  N\ovl{\left(\frac{2f'(0)}{f(0)}\EE -\SS\right)^2} & = & \frac{4}{f(0)} 
  (f f'' -f'^2) + \frac{8}{N} f''(0); \nonumber\\ 
  & = & \frac{8}{N} f''(0).
\end{eqnarray}
is 0 at the order $N^{-1}$. This means that, while $b$ is strictly
positive, there is no possible fluctuations of $\DD$  around
$2f'(0)/f''(0)\times\EE$. 

Repeating the argument for (\ref{eq:npoints:annealed}), $n$ finite,
shows that $\Sp(\HH)$ depends only on $V(\rr)$ and not on its local
environment. Thus, we argue that the
average~(\ref{eq:slice-nm-definition}) is given by
(\ref{eq:width:hessian}, \ref{eq:shift:hessian}), as announced.


\section{The quasi-static picture}
\label{app:three}

From equation~(\ref{eq:r-g}), we derive the expression for
$r(t'+\tau,t')$. 
\begin{equation}
r(t'+\tau,t') = 2 \frac{e^{\SS\tau}\ \mbox{I}_1(\Lambda\tau)\
e^{-\Lambda\tau}}{\Lambda\tau},
\end{equation}
where $\mbox{I}_1$ is the first kind modified Bessel function. 
The short time expansion of (\ref{eq:partial-b-g}) is:
\begin{eqnarray}
\partial_{t'} b(t'+\tau,t') & = & -2\tau \int\dd\lambda\ \rho(\lambda)
g(\lambda,t'); \nonumber\\
  & =  &  -2\tau \Big(\dot{\EE}(t')\Big)^2; \nonumber\\
  & =  &  -2\tau\; \Gamma S^{3/2}.
\end{eqnarray}

In the intermediate time separation regime, $\tau$ is of order
$\SS^{-1}$. The integral is dominated by $\lambda \sim \SS$ and cut off
by $g(\lambda,t')$ for $\lambda\SS \gg 1$. $\rho(\lambda)$ can be
replaced by its $\lambda\to 0$ equivalent.
\begin{eqnarray}
r(t'+\tau,t') & \simeq  & \sqrt{2/\pi} e^{\SS\tau} (\Lambda \tau)^{-3/2};
\nonumber\\
  & = &  \SS^{3/2} (\sqrt{2/\pi} \Lambda^{-3/2})\cdot(\SS \tau)^{-3/2}
  e^{\SS\tau}; \nonumber\\
  & = & \SS^{3/2} \Phi_0(\SS\tau).\\
\partial_{t'} b(t'+\tau,t') & = & -2\Gamma \SS^{1/2}
\int_0^{2\Lambda/\SS \simeq \infty} \dd u\ (2/\Lambda)^{3/2}\pi^{-1}
\sqrt{u}\nonumber\\ 
  &   & \times\hat{G}(u) \left( \frac{1-e^{-\SS\tau(u-1)}}{u-1}
  \right); \nonumber\\ 
  & = & -2 \Gamma S^{1/2} \Phi_1(\SS \tau).
\end{eqnarray}

The effective temperature behaves as:
\begin{equation}
  \Tb(t'+\tau,t') = \left(\frac{\Gamma}{\SS}\right)
  \frac{\Phi_1(\SS\tau)}{\Phi_0(\SS\tau)},
\end{equation}
reducing to 
\begin{equation}
  \Tb(t,t') =  \frac{\Phi_1}{\Phi_0}(\SS\cdot(t-t')),
\end{equation}
if $\kappa=2/3$ and $\Gamma=\SS$.



\clearpage
\centerline{\Large Captions}
\bigskip

FIGURE~1. Parametric plot of the integrated response $b(t,t')$ vs
    ${\mathcal{R}}(t)=\int_{0}^t \dd s\ r(t,s)$ at zero temperature,
    for time steps $h=0.025, 0.5, 1.0, 2.0$. The horizontal part
    corresponds to the short time regime, with $T \to 0$. Then, the
    aging regime is the straight line with a slope $X^{-1} \simeq
    0.21$, to be compared with the theoretical value
    $1/X_{QFD}=2$. The inset shows the derivative $X^{-1}(b) = \dd b/
    \dd {\mathcal{R}}(t)$, stepping from 0 to 2.1. 

FIGURE~2. Dynamical energy ${\cal E}(t)+2$ vs time, in log-log
    coordinate, for $h=0.025$ and $h=0.2$.  The power-law decay is
    unambiguous, and a fit to $\kappa=0.67$ has been done between the
    two vertical arrows.

FIGURE~3. Logarithmic derivative $-\kappa$ of the energy ${\cal
    E}(t)+2$, for $h=0.025, 0.5, 1.0, 2.0$. The curve is noisy as
    ${\cal E}(t) \to -2$. The straight line stands for $ \kappa=2/3$
    which we believe to be its exact value. Curves for $h=0.025, 0.05$
    seems to tend to 2/3 from above, while $h=0.2$ seems to tend to
    2/3 from below. $\kappa \simeq 2/3$ is well realised for $h=1$.

FIGURE~4. The characteristic times $t_f(t) \propto t^{\alpha}$ and
    $t_a(t),\ a=0.55$ and $a=0.45$ determined from
    equations~(\ref{eq:tf}), (\ref{eq:tf-prime}) and
    (\ref{eq:ta}). Results are shown for the time steps $h=0.1$ and
    $h=0.2$, and the finiteness of $h$ is visible at small $t$. A
    numerical estimate of $\alpha$ is $0.64$ between the first and
    last vertical arrows. The exponent of $t_{a=0.55}$ is close to
    $0.93$ while we expect 1, and $t_{a=0.45}$ should saturate to a
    constant.

FIGURE~5. The functions $\exp[-b(t,0)]$, $\exp[-b(t,t-t_f(t))]$ vs
    $t$, for $h=0.2$ and $h=0.1$, in logarithmic coordinates.  The
    functions $\exp[-b(t,20)]$ vs $t-20$ and $\exp[-b(t,40)]$ vs
    $t-40$. Here, 20 and 40 are waiting times.  The behaviour of
    $b(t,0)$ and $b(t,t-t_f(t))$ is doubtless logarithmic.  The slopes
    of $\exp[-b(t,0)]$, $\exp[-b(t,t-t_f(t))]$, on this figure are
    respectively $-1.10$ and $-0.42$.  According to the predictions
    of~\protect\cite{CugLeD}, $\exp(-b(t,t'))$ tends to $h(t)-h(t')$
    for $t,t'\gg 1$; $t/t'$ finite. The curves $\exp[-b(t,20)]$ and
    $\exp[-b(t,40)]$ tend to imitate $\exp[-b(t,0)]$, with a delay.

FIGURE~6. A test of the linear response of the displacement $u(t)$. We
    plot $u(t)/F$, as a function of $\ln(t)$, for
    $F=0.05, 0.1, 0.2$ and $0.4$; $h=0.2$. As $F \to 0$, the curves are
    indistinguishable from the integrated response ${\mathcal
    {R}}(t)$.  A departure from the straight line signals the breakdown of the
    linear response, as the particle  acquires a finite velocity, dependent
    (non-linearly) on the force. The suggested behaviour of $u(t)$ is
    thus: $u(t)=  F\cdot (c_3+c_4\ln(t))$.

FIGURE~7. The density $\rho(\lambda)$ of eigenvalues $\lambda-\SS$ (on
    top).  A sketch of the self-similar distribution $g= \SS
    \hat{G}(\lambda/\SS)$, assuming $\Gamma=\SS$ (bottom). The tail of
    $\hat{G}$ goes to 0 as $\lambda/\SS \to \infty$ , in the
    asymptotic limit $\SS \to 0$.


\begin{figure}
    \epsfxsize 12cm  \centerline{
    \epsfbox{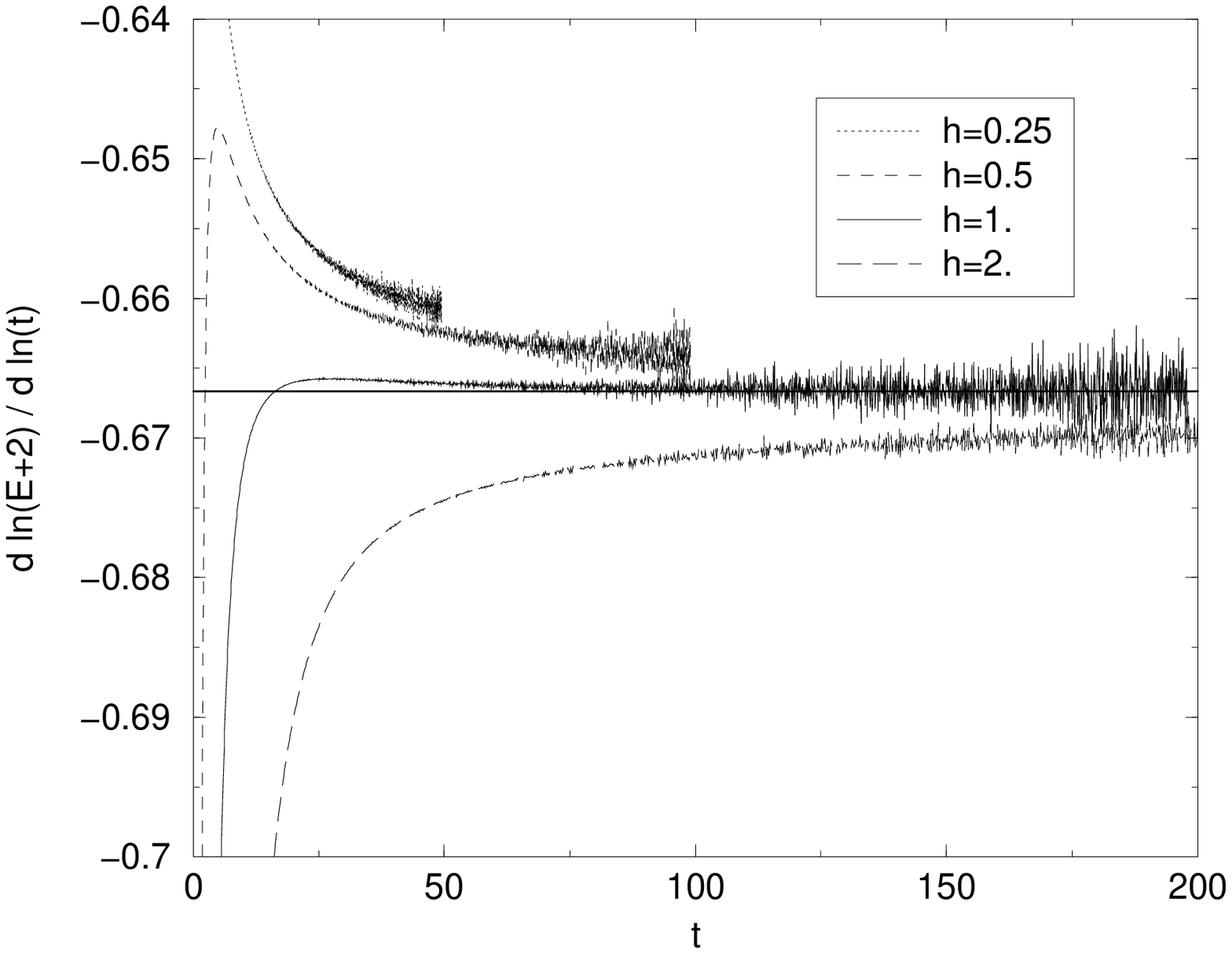} } 
    \caption{ }
    \label{fig:dlog-energy}
\end{figure}


\begin{figure}
    \epsfxsize 12cm  \centerline{
    \epsfbox{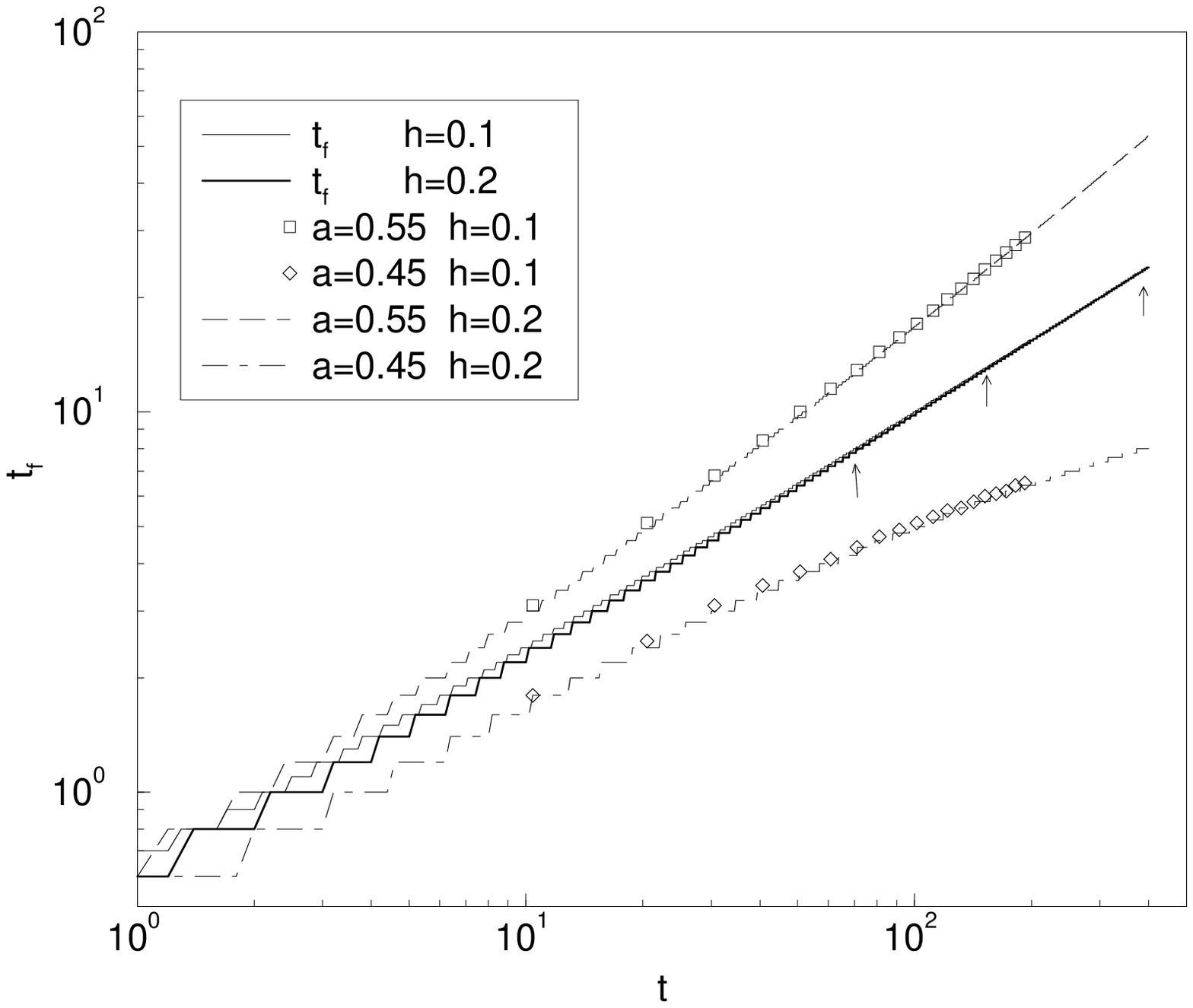} } 
    \caption{ }
    \label{fig:tf}
\end{figure}



\begin{figure}
    \epsfxsize 12cm  \centerline{
    \epsfbox{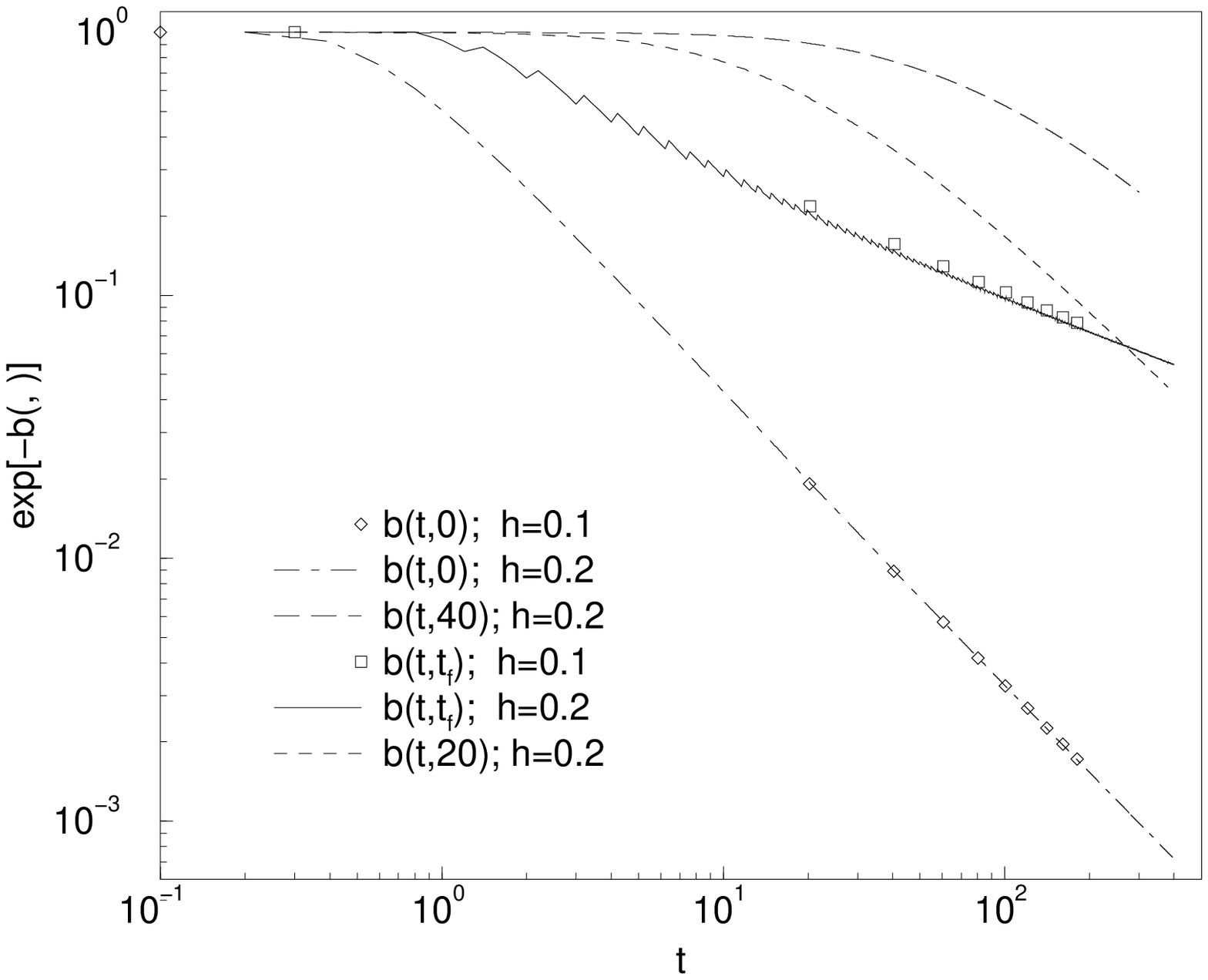} } 
    \caption{ }
    \label{fig:exp-b}
\end{figure}


\begin{figure}
    \epsfxsize 12cm  \centerline{
    \epsfbox{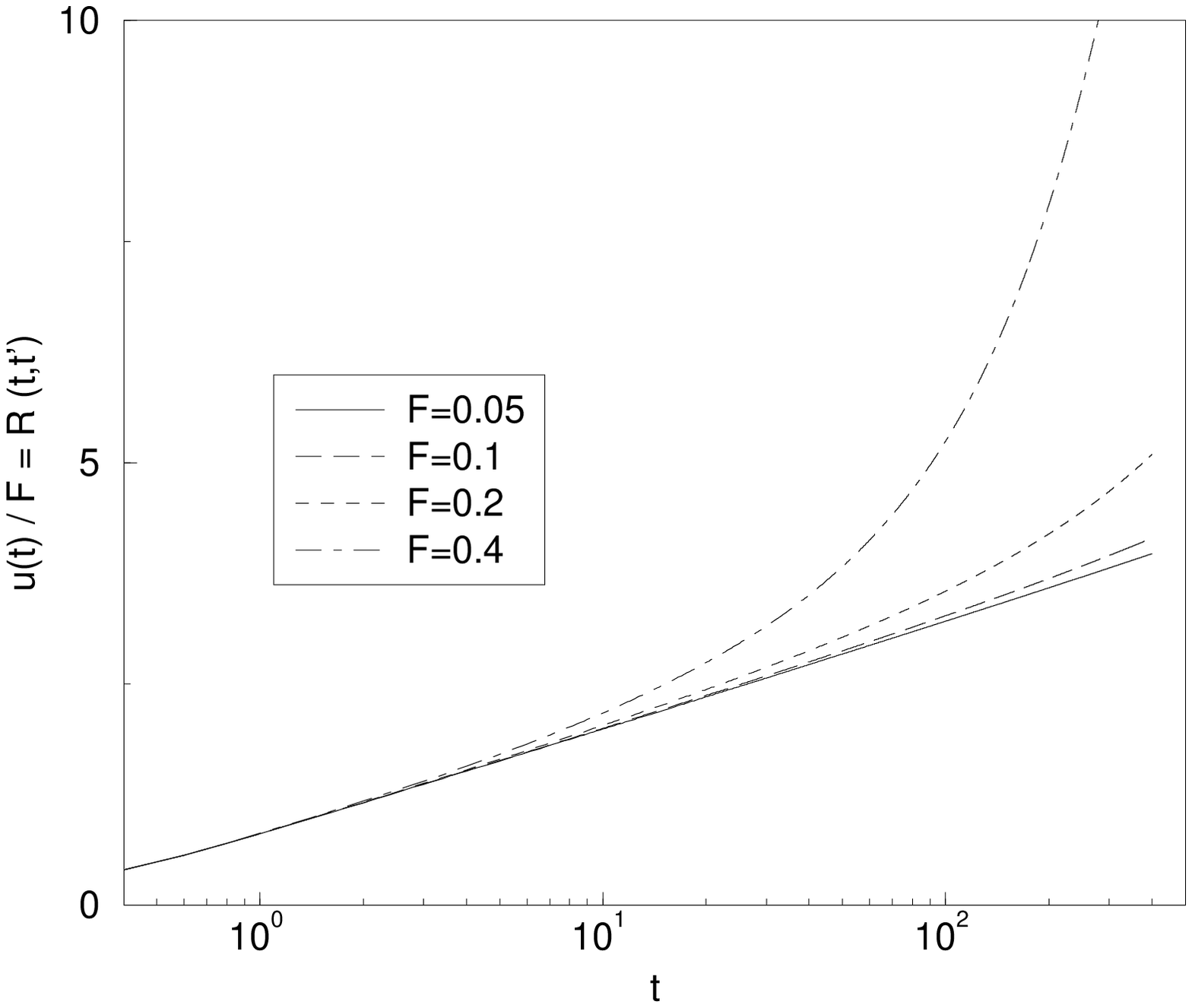} } 
    \caption{ }
    \label{fig:log-disp}
\end{figure}


\begin{figure}
    \epsfxsize 10cm  \centerline{
    \epsfbox{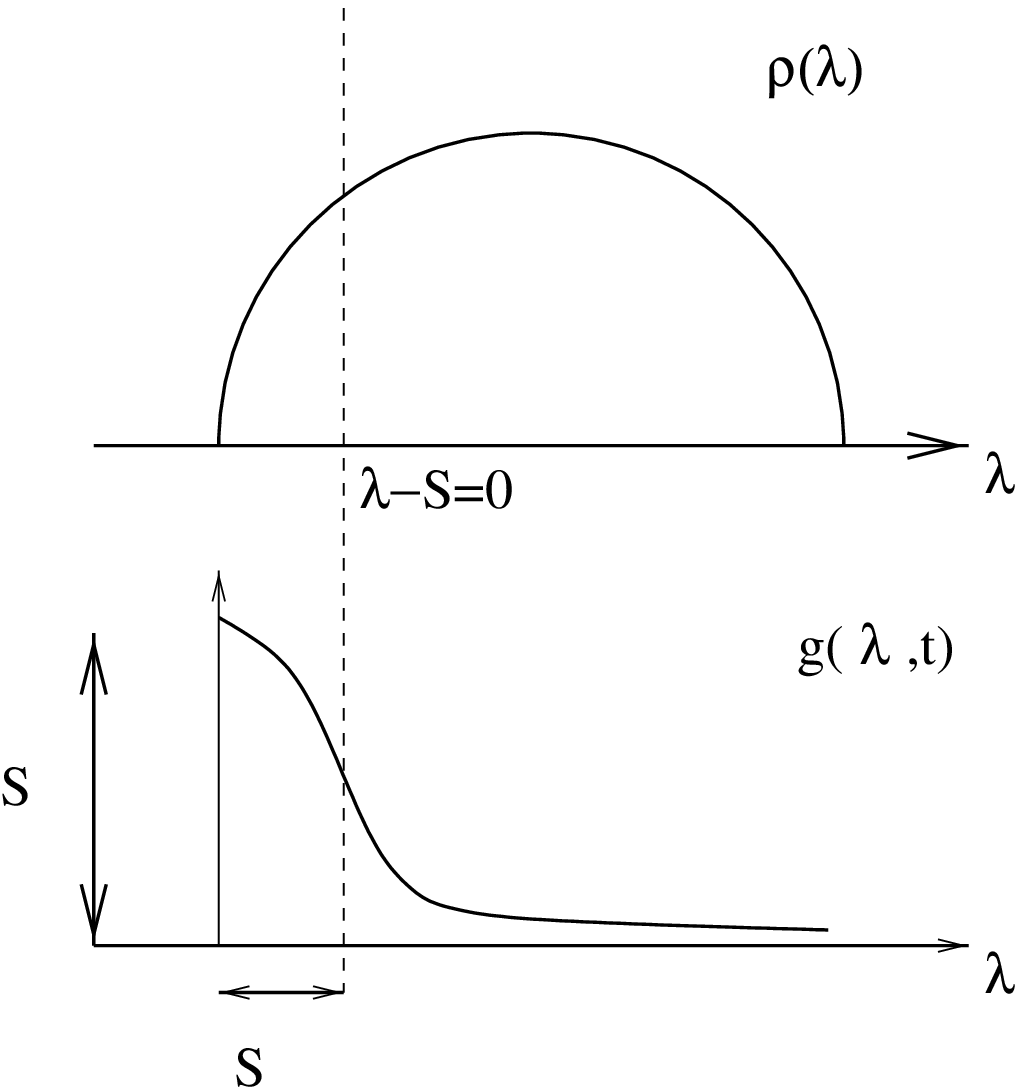} } 
    \caption{ }
    \label{fig:gradient}
\end{figure}


\begin{thebibliography}{10}
\bibitem{rem:MCT} We refer here to the mode coupling equations for
supercooled liquids, as reviewed in~\cite{Gotze}.
\end{thebibliography}
\end{document}